\documentclass{article}
\usepackage{arxiv}

\usepackage[utf8]{inputenc}
\usepackage[T1]{fontenc}
\usepackage{amsmath,amsfonts,amssymb}
\usepackage{array}
\usepackage{booktabs}
\usepackage{tabularx}
\usepackage{textcomp}
\usepackage{url}
\usepackage{verbatim}
\usepackage{graphicx}
\usepackage{placeins}
\usepackage{float}
\usepackage{algorithm}
\usepackage{algorithmicx}
\usepackage{algpseudocode}
\usepackage{microtype}
\usepackage[colorlinks=true,citecolor=blue,linkcolor=blue]{hyperref}

\graphicspath{{figures/}}
\begin{document}

\title{Efficient Multimodal Beam Prediction for Real-World 6G UAV Communications via Image–Position–Synthetic Fusion}

\author{Rifat Bin Reza\thanks{R. B. Reza is with the Department of Electrical and Electronic Engineering, Brac University, Dhaka, Bangladesh. E-mail: \texttt{rifat.bin.reza1@g.bracu.ac.bd}.}, Md Ahadul Haque Suvo\thanks{M. A. H. Suvo is with the Department of Avionics Engineering, Aviation and Aerospace University, Dhaka, Bangladesh. E-mail: \texttt{suvo\_23024006@aaub.edu.bd}.},\\
\textbf{Jia Uddin}\thanks{J. Uddin is with the Department of AI and Big Data, Woosong University, Daejeon, Republic of Korea. E-mail: \texttt{jia.uddin@wsu.ac.kr}.}, and \textbf{Saifur Rahman Sabuj}\thanks{S. R. Sabuj is with the Department of Electrical and Electronic Engineering, Brac University, Dhaka, Bangladesh, and the Department of Automotive Engineering, Hanyang University, Seoul, South Korea. E-mail: \texttt{s.r.sabuj@ieee.org}.}
}

\maketitle

\begin{abstract}
Camera-based beam selection can reduce the need for exhaustive channel measurement in highly mobile millimeter-wave UAV links, but the complementary value of visual and position-derived representations and their behavior under modality degradation require further investigation. This paper develops a staged multimodal framework for current-beam prediction using real-world UAV measurements. The proposed 32-beam classifier combines an ImageNet-pretrained EfficientNet-B3 visual encoder with two deterministic position-derived representations: an explicit Geometric--Kinematic vector and a nonlinear synthetic feature basis constructed from position. The image and synthetic representations are fused first, followed by late fusion with the geometry representation. Experiments on DeepSense~6G Scenario~23 show that Image--Synthetic fusion achieves 86.92\% Top-1, 99.03\% Top-3, and 99.56\% Top-5 accuracy, improving Top-1 performance by 2.64 percentage points over image-only inference and outperforming the 60.71\% Top-1 accuracy of synthetic-only inference. Adding the Geometric--Kinematic branch yields the highest measured performance, reaching 87.14\% Top-1, 99.12\% Top-3, and 99.60\% Top-5 accuracy. Controlled input interventions further demonstrate complementary behavior under visual degradation and strong sensitivity to severe synthetic-input corruption. Position-derived nonlinear features provide complementary information to visual context and improve current-beam prediction in the evaluated real-world UAV setting. 
\end{abstract}

\keywords{Beam prediction \and EfficientNet-B3 \and Millimeter-wave \and Multimodal learning \and UAV communication}

\section{Introduction}
Sixth-generation (6G) wireless systems are expected to support highly mobile platforms such as uncrewed aerial vehicles (UAVs), autonomous transportation systems, aerial sensing networks, and integrated sensing-and-communication services \cite{sabuj2023cuav}. Millimeter-wave (mmWave) and sub-terahertz links can provide the bandwidth required by these applications, but their high propagation loss necessitates large antenna arrays and narrow directional beams. In a UAV link, three-dimensional motion, changing orientation, and rapid variations in range and elevation force the serving beam to be updated frequently. Exhaustive codebook sweep consequently consumes radio resources, increases access latency, and shortens the time available for payload transmission. The first real-world position- and camera-aided UAV study established sensing-assisted learning as a practical route to lowering this overhead \cite{charan2022drone}.

Progress in this area has been accelerated by DeepSense~6G, which provides real-world communication and sensing measurements aligned with time, including mmWave received power, RGB images, GPS, LiDAR, and radar in diverse deployment scenarios \cite{alkhateeb2023deepsense}. Its UAV-focused Scenario~23 contains coexisting camera observations, practical position information, and beam-sweep measurements. These data permit beam selection to be formulated as a supervised classification: environmental and geometric observations are mapped to the index of the beam that maximizes received power. They also expose effects that are difficult to reproduce faithfully with ray-tracing alone, including sensor noise, calibration error, background clutter, and irregular flight trajectories.

Position is an attractive source of information on the side because GPS is already available on most aerial platforms. Morais \emph{et al.} quantified the utility and limitations of practical GPS for real-world beam prediction and showed that localization quality and evaluation criteria materially affect the inferred communication benefit \cite{morais2023position}. Charan \emph{et al.} subsequently fused vision and position on a real mmWave dataset, demonstrating that the two modalities encode complementary scene and user-location information \cite{charan2022multimodal}. Nevertheless, coordinate-only models cannot directly observe blockage or background geometry, and narrow codebooks can turn even a small localization error into an adjacent-beam classification error.

Camera sensing supplies richer environmental context without requiring active ranging hardware. Jiang and Alkhateeb used visual sequences to track future beams in a real mmWave deployment, showing that RGB observations can sharply reduce conventional beam-training overhead \cite{jiang2022vision}. For UAV communication, Ahmad \emph{et al.} combined UAV localization from images with position models and a stacking-based meta-learner \cite{ahmad2023vision}. These studies motivate strong visual encoders, but they also reveal practical challenges: the aerial terminal may occupy only a small image region, illumination and viewpoint vary, and an image-only network must infer metric geometry indirectly from pixels. Other sensing modalities offer complementary tradeoffs. LiDAR-aided prediction captures the explicit three-dimensional structure and has supported the prediction of current and future beams in real vehicle-to-infrastructure links \cite{jiang2023lidar}. Radar-aided tracking provides range, angle, and Doppler cues and has been demonstrated in real vehicle-to-vehicle mmWave experiments \cite{luo2023radar}.

\begin{figure}[!t]
    \centering
    \includegraphics[width=.5\textwidth]{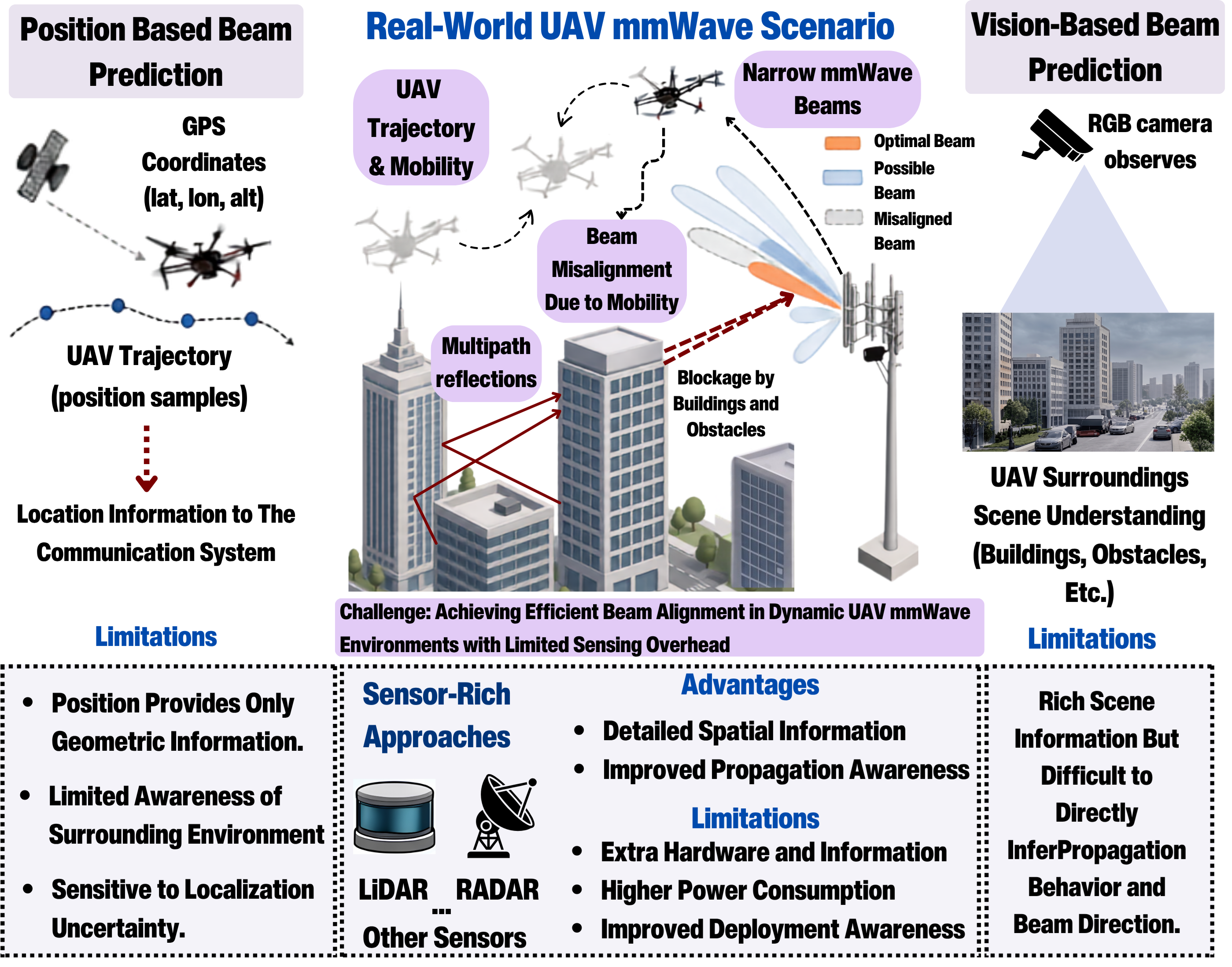}
    \caption{Real-world UAV mmWave beam-selection scenario and sensing tradeoffs. The link must maintain narrow-beam alignment under UAV mobility, blockage, and multipath, while position-only, vision-based, and sensor-rich approaches provide different information-cost profiles.}
    \label{fig:uav_mmwave_scenario}
\end{figure}

Figure~\ref{fig:uav_mmwave_scenario} makes the sensing tradeoff explicit. LiDAR and radar can provide direct range, angle, or structural information, but their mass, power, computation, and integration costs may be difficult to justify in payload-constrained UAVs. Position is inexpensive and already available, yet it does not encode blockage or scene appearance; RGB imagery supplies that context, but only through an indirect, data-driven mapping from pixels to beam state. We therefore combine camera and position observations without introducing an additional active sensor. The synthetic-data lifting does not create new physical information: it re-expresses the position record in a nonlinear basis whose distance- and phase-like coordinates provide an inductive bias for beam classification.

Existing learning-based beam-management studies address three related but distinct sources of side information. Temporal and cross-band methods exploit signal history, dual-band observations, or recurrent dynamics to improve tracking and adaptation \cite{ma2023dualband,khan2024lowcomplexity,li2023timedomain}; their assumptions differ from the current-beam task, which uses one synchronized camera--position observation at inference time.

A second line of work increases environmental observability through radar, LiDAR, semantic context, digital twins, transformers, or multi-sensor fusion \cite{demirhan2022radarbeam,tian2023transformers,
imran2023semantic,shi2024multimodal}. UAV-specific visual backbones, semantic representations, and multi-candidate camera formulations further show that scene context and target ambiguity materially affect learned beam decisions \cite{zarei2023seresnet,yang2023semantics,charan2025multicandidate}. These approaches can improve observability, but they also change the sensing, payload, or supervision budget; the relevant comparison is therefore not accuracy alone, but accuracy conditioned on what information must be acquired at inference time.

A third line of evidence concerns generalization and reliability. Cross-distribution benchmarks, adversarial studies, and recent surveys show that nominal fixed-split accuracy does not by itself characterize robustness \cite{charan2022challenge,catak2022security,xue2024survey}. Real-data position and GPS studies continue to demonstrate the utility of location assistance \cite{gouda2025position,nugroho2025gps}, and extended sensing-aided UAV experiments confirm the value of camera and GPS observations for current and future beam alignment \cite{charan2025sensing}. They also motivate explicit separation of nominal predictive gain, distribution shift, and failure under degraded inputs.

Table~\ref{tab:sota_comparison} positions the present study relative to the five most closely aligned real-data approaches. The unresolved question is not whether position or vision can predict beams in isolation, but whether a deterministic nonlinear reparameterization of already available position contributes information that a strong visual encoder does not readily recover, and whether a separate motion-aware geometry branch remains nonredundant after that lifting has been fused with imagery. A second unresolved issue is reliability: a gain measured on one fixed trajectory is insufficient unless the predictor is also examined under modality degradation. We address these questions with a nested architecture in which an EfficientNet-B3 image representation is fused first with a compact synthetic-data basis and then, diagnostically, with explicit geometry and motion. This construction supports controlled comparisons among image-only, synthetic-only, image--synthetic, and full-fusion hypotheses while preserving the distinction between a learned representation of position and an independently measured sensing modality. 

\begin{table}[!t]
\caption{Comparison of State-of-the-Art Sensing-Aided Beam Prediction Models}
\label{tab:sota_comparison}

\centering
\footnotesize
\setlength{\tabcolsep}{2.5pt}
\renewcommand{\arraystretch}{1.2}

\begin{tabularx}{\columnwidth}{|l|>{\raggedright\arraybackslash}X|>{\raggedright\arraybackslash}X|>{\raggedright\arraybackslash}X|}
\hline
\textbf{Reference} & \textbf{Baseline Approach} & \textbf{Limitation} & \textbf{Proposed Upgrade} \\
\hline

Jiang \cite{jiang2022vision} & Visual seq-to-seq tracking & Tracking-focused; not designed for current UAV beam classification & Image--synthetic fusion for current-beam inference \\
\hline

Charan \cite{charan2022drone} & ResNet-50 vision or position-based model & Separate branches; lacks synthetic modeling  &
EfficientNet-B3, synthetic lifting \\
\hline

Ahmad \cite{ahmad2023vision} & YOLO-based vision with position stacking & Depends on detected bounding boxes &
Learns full-frame visual and synthetic representations \\
\hline

Charan \cite{charan2025sensing} & Sensing-aided tracking framework & No explicit synthetic transformation &
Introduces synthetic lifting with controlled input-intervention auditing \\
\hline

Zarei \cite{zarei2025lstm} & LSTM-based image or GPS prediction & Single-modality future-step prediction &
Multimodal late fusion for current-beam prediction \\
\hline
\end{tabularx}
\end{table}

The main contributions are summarized as follows:
\begin{enumerate}
    \item We develop a staged EfficientNet-B3 beam-prediction architecture for DeepSense~6G Scenario~23 that sequentially learns visual and position-derived representations, performs image--synthetic fusion, and incorporates explicit Geometric--Kinematic features through late fusion.
    \item We formulate and explicitly characterize a deterministic nonlinear position-derived feature transformation, clearly distinguishing this synthetic representation from measured channel state information (CSI).
    \item We conduct a controlled ablation of synthetic-only, image-only, image--synthetic, and full-fusion configurations on a common fixed evaluation partition, quantifying the contribution of each representation through Top-$k$ beam-selection accuracy.
    \item We evaluate multimodal reliability through controlled test-time input interventions and complement the stress analysis with performance-based group attribution, revealing the complementary value of position-derived features under nominal conditions and the sensitivity of static fusion to severe modality degradation.
\end{enumerate}

Table~\ref{tab:notation} summarizes the principal notation used throughout the system model and learning formulation.
\begin{table}[!t]
\caption{Principal Notation}
\label{tab:notation}
\centering
\footnotesize
\setlength{\tabcolsep}{3pt}
\renewcommand{\arraystretch}{1.08}
\begin{tabularx}{\columnwidth}{>{\centering\arraybackslash}p{0.23\columnwidth} X}
\toprule
\textbf{Symbol} & \textbf{Definition} \\
\midrule
$n,k,q$ &
Sample, OFDM-subcarrier, and retained-beam indices \\
$M,K,Q_0,Q$ &
Number of antenna elements, subcarriers, original and retained beams \\
$\mathbf{b}^{(0)}_p,\mathbf{b}_q$ &
Original and retained receive-combining codewords \\
$\mathbf{h}_{n,k}$ &
Uplink channel vector \\
$\mathcal{P}_{n,q}$ &
Measured received power of retained beam $q$ \\
$q_n^\star$ &
Ground-truth optimal retained-beam index \\
$\mathbf{I}_n$ &
Synchronized RGB image \\
$(x_n,y_n)$ &
Planar UAV Cartesian coordinates \\
$r_n,\theta_n$ &
Planar range and azimuth \\
$\Delta x_n,\Delta y_n$ &
Consecutive planar position increments \\
$\delta_n,\psi_n$ &
Inter-sample planar displacement magnitude and motion heading \\
$\mathbf{d}_n\in\mathbb{R}^{11}$ &
Position-derived synthetic feature vector \\
$\mathbf{g}_n\in\mathbb{R}^{15}$ &
Geometric--kinematic feature vector \\
$\phi_{\mathrm{EffB3}}(\cdot)$ &
Pretrained EfficientNet-B3 image backbone \\
$\mathbf{z}_{n,I},\mathbf{z}_{n,D},\mathbf{z}_{n,G}$ &
Image, synthetic-data, and geometry branch embeddings ($\in\mathbb{R}^{512},\mathbb{R}^{256},\mathbb{R}^{96}$) \\
$\mathbf{z}_{n,ID}$ &
Fused image--synthetic embedding, $\in\mathbb{R}^{256}$ \\
$\mathbf{o}_n^{(IDG)}$ &
Final $Q$-dimensional beam-logit vector \\
$\pi_{n,q}$ &
Posterior probability assigned to beam $q$ \\
$\hat q_n$ &
Predicted beam index \\
$\boldsymbol{\Theta}_I, \boldsymbol{\Theta}_D,$ \newline $\boldsymbol{\Theta}_{ID}, \boldsymbol{\Theta}_{IDG}$ & Stage-specific trainable parameter sets \\
$\eta_{(\cdot)},T_{(\cdot)},B$ &
Stage-specific learning rate, epoch budget, mini-batch \\
$\epsilon$ &
Label-smoothing factor ($\epsilon=0.03$) \\
$\mathcal{T}_{n,k}$ &
Indices of the $k$ highest-probability beams \\
$\operatorname{Acc}@k,$ &
Top-$k$ accuracy\\
\bottomrule
\end{tabularx}
\end{table}

% =========================================================
\section{System Model and Problem Formulation}
\label{sec:system_model}
% =========================================================
\subsection{Wireless Link and Beam-Selection Model}
Consider a single-user mmWave uplink from a single-antenna UAV transmitter to a next-generation Node B (gNB) receiver equipped with an $M$-element antenna array and a receive-combining codebook. This configuration matches the measurement setup: the UAV carries a quasi-omnidirectional transmit antenna, while the gNB applies a phased-array receive combiner. The gNB is also equipped with an RGB camera, while synchronized UAV position information is available for multimodal beam prediction. The communication waveform
contains $K$ orthogonal frequency-division multiplexing (OFDM) subcarriers. Let the original measured beam-sweep codebook be
\begin{equation}
\mathcal{B}_0=
\left\{
\mathbf{b}^{(0)}_0,
\mathbf{b}^{(0)}_1,
\ldots,
\mathbf{b}^{(0)}_{Q_0-1}
\right\},
\end{equation}
where $\mathbf{b}^{(0)}_p\in\mathbb{C}^{M\times1}$, $\left\|\mathbf{b}^{(0)}_p\right\|_2^2=1$, $p=0,\ldots,Q_0-1$. Here, $M$ denotes the number of gNB antenna elements and $Q_0$ denotes the number of beams in the original measured codebook. The preprocessing retains every second codeword, indexed from $p=0$; this construction requires $Q_0$ to be even, which holds for the retained configuration below. Thus,
\begin{equation}
Q=\frac{Q_0}{2},
\qquad
\mathbf{b}_q=\mathbf{b}^{(0)}_{2q},
\qquad
q=0,\ldots,Q-1.
\end{equation}
For the measurement configuration considered in this work, $Q_0=64$ and hence $Q=32$. For observation $n$ and OFDM subcarrier $k$, let $\mathbf{h}_{n,k}\in\mathbb{C}^{M\times1}$ denote the uplink channel vector after OFDM demodulation. Let $x_{n,k}$ and $\mathbb{E}\!\left[|x_{n,k}|^2\right]=P_{n,k}$ denote the transmitted data symbol and its average power, respectively. When beam $q$ is applied, the received post-demodulation symbol is
\begin{equation}
y_{n,k}^{(q)}
=
\mathbf{b}_{q}^{H}\mathbf{h}_{n,k} x_{n,k}
+
u_{n,k}^{(q)},
\label{eq:received_signal}
\end{equation}
where $\mathbf{b}_q$ denotes the $q$th receive-combining vector applied at the gNB, $x_{n,k}$ is the symbol transmitted by the UAV, $u_{n,k}^{(q)}$ represents aggregate receiver noise and residual interference, including residual inter-carrier and inter-symbol interference when channel time variation prevents an ideal diagonal OFDM representation. Its second moment is $\nu_{n,k,q}^{2} = \mathbb{E} \left[ \left|u_{n,k}^{(q)}\right|^2 \right]$. This quantity equals the disturbance variance under the zero-mean assumption $\mathbb{E}[u_{n,k}^{(q)}]=0$ and additive white Gaussian noise (AWGN) $w_{n,k}\sim\mathcal{CN}(0,\sigma_{n,k}^{2})$ and $u_{n,k}^{(q)}=w_{n,k}$. The corresponding subcarrier SNR is
\begin{equation}
\gamma_{n,k}^{(q)}
=
\frac{
P_{n,k}
\left|
\mathbf{b}_q^{H}\mathbf{h}_{n,k}
\right|^2
}{
\nu_{n,k,q}^{2}
}.
\label{eq:sinr}
\end{equation}

The average beam utility over the OFDM subcarriers is
\begin{equation}
\bar{\gamma}_{n}^{(q)}
=
\frac{1}{K}
\sum_{k=0}^{K-1}
\gamma_{n,k}^{(q)}.
\label{eq:avg_sinr}
\end{equation}

Accordingly, under the arithmetic-mean subcarrier utility in \eqref{eq:avg_sinr} -- one reasonable choice among several possible per-beam utility aggregations, such as a worst-case subcarrier SNR or a rate-weighted average -- the resulting SNR-maximizing beam is
\begin{equation}
q_{n,\mathrm{SNR}}^{\star}
=
\underset{q\in\{0,\ldots,Q-1\}}
{\arg\max}
\;
\bar{\gamma}_{n}^{(q)}.
\label{eq:sinr_best_beam}
\end{equation}

\subsection{Problem Formulation and Fusion Overview}
Explicit CSI is notoriously difficult to acquire in highly mobile mmWave systems, and exhaustive codebook sweeps introduce unacceptable latency. Therefore, rather than measuring the channel directly, we formulate beam selection as a prediction task driven by out-of-band sensory data. We assume the gNB has access to a synchronized multimodal observation tuple comprising three distinct inputs: ($i$) an RGB image taken by the gNB camera, ($ii$) geometric--kinematic telemetry fed back from the UAV, and ($iii$) a deterministic position-derived synthetic feature vector computed from that telemetry. This synthetic vector is a handcrafted feature transformation, not measured CSI and not the output of channel estimation.
 
The supervised target is the empirical receiver-beam label obtained from the measured Scenario~23 beam-sweep powers. Following the retained 32-beam task definition, the label corresponds to the receiver-side beam that maximizes the recorded received power after the dataset preprocessing step. Thus, the learning task is to map the synchronized image, geometric--kinematic, and synthetic inputs to a posterior ranking over the retained receiver-beam classes. The highest-scoring class is used as the predicted beam, while the full ranking supports Top-$k$ candidate evaluation.
 
The proposed network follows a staged fusion strategy rather than directly concatenating all raw modalities. The synthetic branch first learns a compact representation $\mathbf{z}_{n,D}$ from the position-derived feature vector $\mathbf{d}_n$. In parallel, the RGB encoder extracts a visual representation $\mathbf{z}_{n,I}$. These two representations are subsequently combined by an image--synthetic fusion module to obtain the intermediate embedding $\mathbf{z}_{n,ID}$. A separate geometric encoder transforms the explicit geometric--kinematic vector $\mathbf{g}_n$ into $\mathbf{z}_{n,G}$. Finally, $\mathbf{z}_{n,ID}$ and $\mathbf{z}_{n,G}$ are fused through a geometry-guided late-fusion classifier that produces the posterior probabilities of the retained beam classes.
 
This staged design separates representation learning from final geometry-guided refinement. The image branch contributes scene and target-location cues that cannot be recovered from coordinates alone; the synthetic branch provides a structured nonlinear lifting of the position record without introducing additional physical sensing; and the geometric--kinematic branch keeps the original spatial and motion descriptors available for final refinement. This preserves the measurement-consistent uplink interpretation of Scenario~23 established in Section~\ref{sec:system_model} while avoiding any assumption of instantaneous CSI or downlink--uplink reciprocity.

% =========================================================
\subsection{Beam-Label Construction and Learning Problem}
% =========================================================

The experimental dataset does not require explicit instantaneous CSI for constructing the supervised targets. Instead, the beam label is obtained directly from the measured beam-sweep powers. Let $\mathcal{P}_{n,p}^{(0)}$ denote the measured received power associated with the $p$th beam of the original codebook at observation $n$. After retaining every second beam, $\mathcal{P}_{n,q} = \mathcal{P}_{n,2q}^{(0)}, ~ q=0,\ldots,Q-1$. The supervised ground-truth beam label is therefore defined as
\begin{equation}
q_n^\star
=
\underset{q\in\{0,\ldots,Q-1\}}
{\arg\max}
\;
\mathcal{P}_{n,q}.
\label{eq:measured_best_beam}
\end{equation}

When the effective noise and residual-interference levels are approximately invariant across the swept beams, maximizing the measured received power is consistent with maximizing the corresponding SINR; this equivalence additionally presumes consistent transmit power, consistent aggregation across OFDM subcarriers, and consistent linear-power processing for every beam in the sweep. Under these conditions, $q_n^\star$ is used as the target class for supervised learning. $\mathcal{P}_{n,q}$ is the linear received power averaged over the beam-sweep measurement window, as provided in the DeepSense~6G Scenario~23 release; ties are resolved by the dataset's native ordering.  

Let $\mathbf{I}[t]\in\mathbb{R}^{W\times H\times C}$ denote the RGB frame with width $W$, height $H$, and $C$ color channels. Let $\mathbf{g}[t] \in \mathbb{R}^{15}$ and $\mathbf{d}[t] \in \mathbb{R}^{11}$ denote the geometric--kinematic and deterministic position-derived synthetic feature vectors, respectively. The joint observation at time $t$ is thus $\mathcal{S}[t] = \{\mathbf{I}[t], \mathbf{g}[t], \mathbf{d}[t]\}$. For each sample $n$, the multimodal observation $\mathcal{S}_n$ is
\begin{equation}
\mathcal{S}_n
=
\left(
\mathbf{I}_n,
\mathbf{g}_n,
\mathbf{d}_n
\right),
\label{eq:multimodal_input}
\end{equation}
The multimodal classifier maps each observation to $Q$ beam-class logits:
\begin{equation}
f_{\boldsymbol{\Theta}}:
\mathcal{S}_n
\rightarrow
\mathbb{R}^{Q},
\label{eq:classifier}
\end{equation}
where $\boldsymbol{\Theta}$ denotes the trainable network parameters. Let $\mathbf{o}_n = [o_{n,0},\ldots,o_{n,Q-1}]^T$ denote the output-logit vector. The posterior probability assigned to beam $q$ is $\pi_{n,q} = \frac{\exp(o_{n,q})}
{\sum_{\ell=0}^{Q-1}\exp(o_{n,\ell})}$. The predicted beam index is
\begin{equation}
\widehat q_n
=
\underset{q\in\{0,\ldots,Q-1\}}
{\arg\max}
\;
\pi_{n,q}.
\label{eq:prediction}
\end{equation}

For a training set containing $N_{\mathrm{tr}}$ observations, the supervised learning problem is formulated as empirical risk minimization:
\begin{equation}
\boldsymbol{\Theta}^{\star}
=
\underset{\boldsymbol{\Theta}}
{\arg\min}
\;
\frac{1}{N_{\mathrm{tr}}}
\sum_{n=1}^{N_{\mathrm{tr}}}
\mathcal{L}
\left(
f_{\boldsymbol{\Theta}}(\mathcal{S}_n),
q_n^\star
\right).
\label{eq:erm}
\end{equation}

The observations originate from sequential UAV trajectories, and the Geometric--Kinematic representation explicitly uses adjacent samples. Consequently, temporally neighboring records can share similar positions and visual context across the sample-level partition. Section~\ref{subsec:reproducibility} quantifies this temporal proximity and its measured effect on Top-1 accuracy.

% =========================================================
\subsubsection{Geometric--Kinematic Feature Construction}
% =========================================================

The explicit geometry branch captures both instantaneous spatial configuration and local trajectory evolution. Let the planar UAV position relative to the gNB be $\mathbf{p}_n=[x_n,y_n]^T$. The corresponding planar range is $r_n
= \sqrt{x_n^2+y_n^2}$, and the azimuth angle is $\theta_n  = \operatorname{atan2}(y_n,x_n)$. For consecutive indexed observations, define $\Delta x_n=x_n-x_{n-1}$ and $\Delta y_n=y_n-y_{n-1}$. The preprocessing does not divide these increments by the inter-sample time interval. Consequently, the corresponding quantity is interpreted as an inter-sample planar displacement magnitude rather than a physical velocity: $\delta_n = \sqrt{ (\Delta x_n)^2+
(\Delta y_n)^2}$. The associated motion heading is $\psi_n = \operatorname{atan2}(\Delta y_n,\Delta x_n)$.

The resulting 15-dimensional geometric--kinematic vector is
\begin{equation}
\begin{aligned}
\mathbf{g}_n = \big[ & x_n, y_n, r_n, \theta_n, \sin\theta_n, \cos\theta_n, x_n^2, y_n^2, x_ny_n, \\
& \Delta x_n, \Delta y_n, \delta_n, \psi_n, \sin\psi_n, \cos\psi_n \big]^\top.
\end{aligned}
\label{eq:geometry_vector}
\end{equation}

Thus, $\mathbf{g}_n\in\mathbb{R}^{15}$. For the first indexed observation, where no preceding position is available, the
difference-based quantities are initialized consistently with the preprocessing procedure. Each geometric feature is standardized using statistics estimated from the training partition:
\begin{equation}
g_{n,j}^{(\mathrm{std})}
=
\frac{
g_{n,j}-\mu_{g,j}
}{
\sigma_{g,j}
},
\qquad j=1,\ldots,15 .
\label{eq:geometry_standardization}
\end{equation}

The standardized vector is denoted by
$\mathbf{g}_n^{(\mathrm{std})}\in\mathbb{R}^{15}$. Here $\mu_{g,j}$ and $\sigma_{g,j}$ denote the empirical mean and
standard deviation of the $j$th feature.

% =========================================================
\subsubsection{Synthetic Position-Derived Features}
% =========================================================

The synthetic branch constructs deterministic nonlinear transformations of the planar position. These features are
position-derived descriptors and should not be interpreted as directly measured complex CSI. A deterministic complex-valued spatial quantity is first constructed from the UAV Cartesian coordinates as $\widetilde h_n = x_n+j y_n$. 
Its magnitude and phase directly correspond to the radial distance and azimuth: $|\widetilde h_n|=r_n, ~
\angle\widetilde h_n=\theta_n ~ \ell_n=\log(1+r_n)$. A synthetic periodic range feature is defined as $\phi_n =
2\pi \frac{r_n}{r_{\mathrm{ref}}}$, where $r_{\mathrm{ref}}>0$ is a declared range scale selected using training-set statistics only, so that $\phi_n$ is dimensionless. This quantity is a deterministic periodic transformation of range, not a physical propagation phase, and is labeled accordingly throughout.

A range-decay basis feature is defined as $\alpha_n=\left(\frac{r_{\mathrm{ref}}}{r_n+\epsilon_r}\right)^{\gamma}$, where $\epsilon_r>0$ avoids numerical division by zero, $r_{\mathrm{ref}}>0$ is a declared range scale, and $\gamma>0$ is a decay exponent, both selected using training-set statistics only; this construction is dimensionless and is described as a range-decay basis rather than as physical channel attenuation. The complete 11-dimensional position-derived synthetic-data vector is
\begin{equation}
\begin{aligned}
\mathbf{d}_n = \big[ & x_n, y_n, r_n, \theta_n, \sin\theta_n, \cos\theta_n, \\
& \phi_n, \alpha_n, \ell_n, \alpha_n\cos\phi_n, \alpha_n\sin\phi_n \big]^\top \!\!\! \in \mathbb{R}^{11}.
\end{aligned}
\label{eq:synthetic_vector}
\end{equation}

The synthetic features are normalized using training-set statistics. Unlike the Geometric--Kinematic vector, the synthetic features are scaled to a $[0,1]$ range using training-set min--max normalization:
For feature index $j$,
\begin{equation}
d_{n,j}^{(\mathrm{norm})}
=
\frac{
d_{n,j}-d_{j,\min}^{\mathrm{tr}}
}{
d_{j,\max}^{\mathrm{tr}}-d_{j,\min}^{\mathrm{tr}}+\epsilon_d
},
\qquad j=1,\ldots,11 ,
\label{eq:synthetic_normalization}
\end{equation}
where $d_{j,\min}^{\mathrm{tr}}$ and $d_{j,\max}^{\mathrm{tr}}$ denote the minimum and maximum of feature $j$ computed exclusively on the training partition; the signed feature value $d_{n,j}$ is used directly, without an absolute value, so that signed coordinates, angles, and periodic components are preserved rather than folded together, and $\epsilon_d>0$ prevents numerical division by zero.

\subsubsection{RGB Image Feature Construction}

Before passing the RGB frames to the visual encoder, training images undergo stochastic augmentation, including random resized cropping, color perturbation, small-angle rotation, and random erasing. For evaluation, each image is resized to $256\times256$ pixels and center-cropped to $224\times224$ pixels using a deterministic transform. Pixel intensities are normalized to the $[-1,1]$ interval as
\begin{equation}
I_n^{(\mathrm{norm})} = 2\left(\frac{I_n}{255}\right)-1.
\label{eq:image_normalization}
\end{equation}
The evaluation transform is applied identically to all test observations.

% =========================================================
\subsection{Multimodal Architecture}
% =========================================================

The proposed network contains three representation branches: an image branch, a synthetic position-derived branch, and a geometric--kinematic branch. The network is trained in stages and combines the three learned representations through an image--synthetic fusion stage followed by a geometry-guided final fusion stage. Let $\operatorname{LN}(\cdot)$ denote layer normalization and $\operatorname{GELU}(\cdot)$ denote the nonlinear activation of the Gaussian error linear unit.

% ---------------------------------------------------------
\subsubsection{Design Overview}
\begin{figure}[!t]
    \centering
    \includegraphics[width=\textwidth]{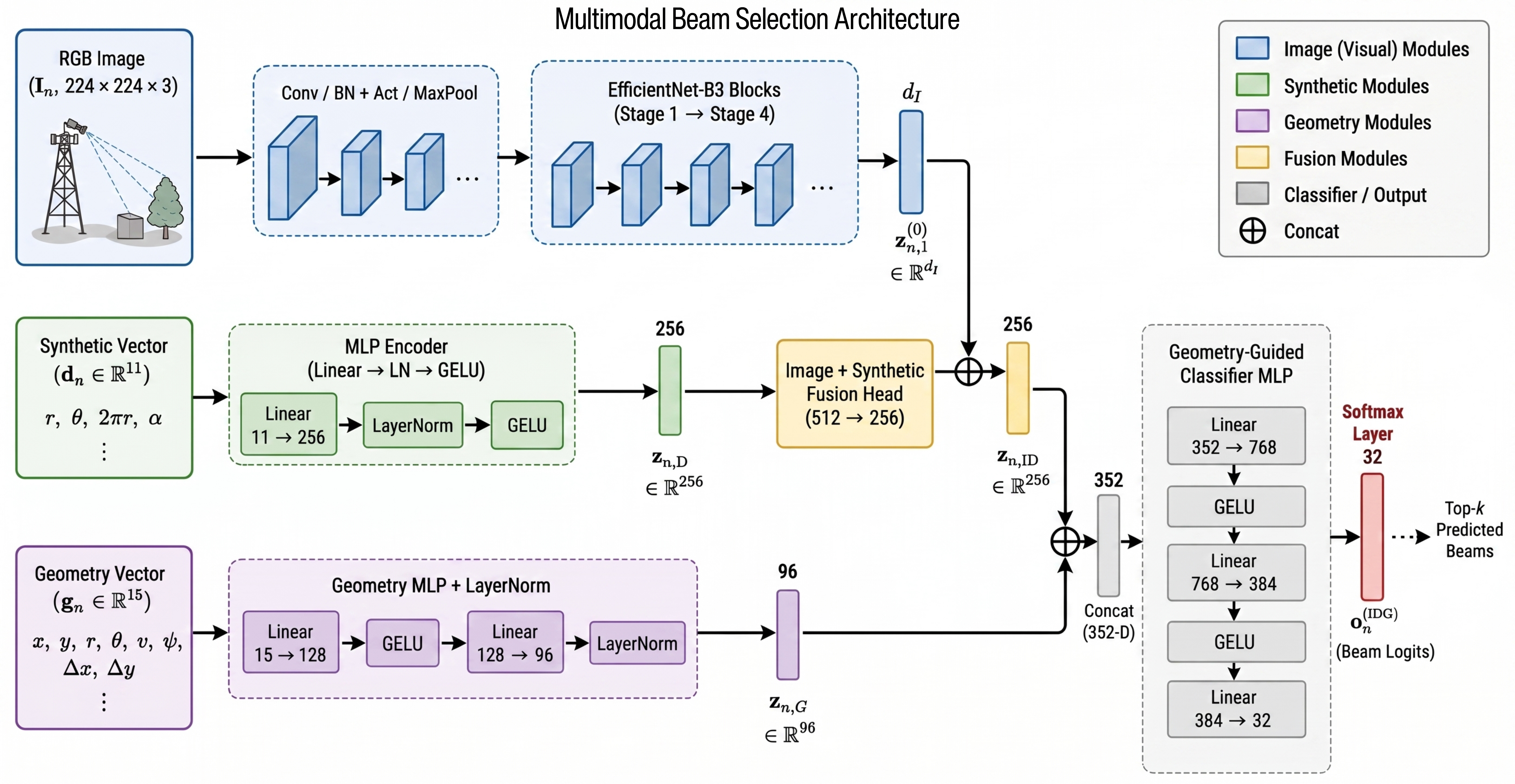}
    \caption{Proposed staged fusion of RGB imagery, geometry-derived synthetic-data features, and explicit Geometric--Kinematic features for 32-beam prediction.}
    \label{fig:architecture}
\end{figure}

% ---------------------------------------------------------
Figure~\ref{fig:architecture} summarizes the proposed network at a high level; the precise input representations, layer dimensions, and staged optimization procedures are formalized in the remainder of this subsection after the input features are defined. The image branch employs an ImageNet-pretrained EfficientNet-B3 backbone \cite{tan2019efficientnet}, the synthetic-data branch encodes the position-derived vector, and the geometric--kinematic branch captures explicit spatial and motion structure. Rather than attempting to learn all representations simultaneously, the modalities are integrated sequentially. The image and synthetic-data representations are fused
first, after which a geometry-guided late-fusion head refines the joint prediction. Because earlier stages are frozen while a later stage is fitted, the incremental contribution of each fusion block remains experimentally interpretable.

% ---------------------------------------------------------
\subsubsection{Image Encoder}
% ---------------------------------------------------------

The synchronized RGB image is processed using an ImageNet-pretrained EfficientNet-B3 backbone:
\begin{equation}
\mathbf{z}_{n,I}^{(0)}
=
\phi_{\mathrm{EffB3}}
\left(
\mathbf{I}_n;\boldsymbol{\Theta}_I
\right)
\in\mathbb{R}^{d_I}.
\label{eq:image_backbone}
\end{equation}

A projection layer with layer normalization and GELU activation produces the visual embedding
\begin{equation}
\mathbf{z}_{n,I}
=
\operatorname{GELU}
\left[
\operatorname{LN}
\left(
\mathbf{W}_{I}\mathbf{z}_{n,I}^{(0)}
+
\mathbf{b}_{I}
\right)
\right]
\in\mathbb{R}^{512}.
\label{eq:image_projection}
\end{equation}

An auxiliary image-only classification head is used during the corresponding pretraining stage and is discarded before multimodal fusion.

% ---------------------------------------------------------
\subsubsection{Synthetic-Feature Encoder}
% ---------------------------------------------------------

The normalized synthetic feature vector is first transformed as
\begin{equation}
\mathbf{u}_{n,D}
=
\operatorname{GELU}
\left[
\operatorname{LN}
\left(
\mathbf{W}_{D,1}\mathbf{d}_n
+
\mathbf{b}_{D,1}
\right)
\right]
\in\mathbb{R}^{256},
\label{eq:synthetic_hidden}
\end{equation}
followed by
\begin{equation}
\mathbf{z}_{n,D}
=
\operatorname{GELU}
\left[
\operatorname{LN}
\left(
\mathbf{W}_{D,2}\mathbf{u}_{n,D}
+
\mathbf{b}_{D,2}
\right)
\right]
\in\mathbb{R}^{256}.
\label{eq:synthetic_embedding}
\end{equation}

The implementation employs fully connected transformations with normalization, GELU activation, and dropout. A temporary auxiliary classification head is used during synthetic-branch pretraining and is removed after pretraining.

% ---------------------------------------------------------
\subsubsection{Geometric--Kinematic Encoder}
% ---------------------------------------------------------

The standardized geometric--kinematic representation is encoded as
\begin{equation}
\mathbf{u}_{n,G}
=
\operatorname{GELU}
\left[
\operatorname{LN}
\left(
\mathbf{W}_{G,1}\mathbf{g}_n
+
\mathbf{b}_{G,1}
\right)
\right]
\in\mathbb{R}^{128},
\label{eq:geometry_hidden}
\end{equation}
and
\begin{equation}
\mathbf{z}_{n,G}
=
\operatorname{GELU}
\left[
\operatorname{LN}
\left(
\mathbf{W}_{G,2}\mathbf{u}_{n,G}
+
\mathbf{b}_{G,2}
\right)
\right]
\in\mathbb{R}^{96}.
\label{eq:geometry_embedding}
\end{equation}

% ---------------------------------------------------------
\subsubsection{Image--Synthetic Fusion}
% ---------------------------------------------------------

The image and synthetic representations are first concatenated:
\begin{equation}
\mathbf{c}_{n,ID}
=
\left[
\mathbf{z}_{n,I};
\mathbf{z}_{n,D}
\right]
\in\mathbb{R}^{768}.
\label{eq:image_synthetic_concat}
\end{equation}

The resulting representation is processed by two fusion layers. The first fusion transformation is
\begin{equation}
\mathbf{u}_{n,ID}^{(1)}
=
\operatorname{GELU}
\left[
\operatorname{LN}
\left(
\mathbf{W}_{ID,1}\mathbf{c}_{n,ID}
+
\mathbf{b}_{ID,1}
\right)
\right],
\label{eq:id_fusion_1}
\end{equation}
followed by
\begin{equation}
\mathbf{z}_{n,ID}
=
\operatorname{GELU}
\left[
\operatorname{LN}
\left(
\mathbf{W}_{ID,2}\mathbf{u}_{n,ID}^{(1)}
+
\mathbf{b}_{ID,2}
\right)
\right]
\in\mathbb{R}^{256}.
\label{eq:id_fusion_2}
\end{equation}

A temporary image--synthetic classification head is used during the corresponding fusion-pretraining stage and is removed before the final geometry-guided stage.

% ---------------------------------------------------------
\subsubsection{Final Geometry-Guided Fusion}
% ---------------------------------------------------------

The intermediate image--synthetic representation is fused with the geometric--kinematic representation:
\begin{equation}
\mathbf{c}_{n,IDG}
=
\left[
\mathbf{z}_{n,ID};
\mathbf{z}_{n,G}
\right]
\in\mathbb{R}^{352}.
\label{eq:idg_concat}
\end{equation}

The first geometry-guided fusion layer is
\begin{equation}
\mathbf{u}_{n,IDG}^{(1)}
=
\operatorname{GELU}
\left[
\operatorname{LN}
\left(
\mathbf{W}_{IDG,1}\mathbf{c}_{n,IDG}
+
\mathbf{b}_{IDG,1}
\right)
\right],
\label{eq:idg_fusion_1}
\end{equation}
followed by
\begin{equation}
\mathbf{u}_{n,IDG}^{(2)}
=
\operatorname{GELU}
\left[
\operatorname{LN}
\left(
\mathbf{W}_{IDG,2}\mathbf{u}_{n,IDG}^{(1)}
+
\mathbf{b}_{IDG,2}
\right)
\right].
\label{eq:idg_fusion_2}
\end{equation}

The final $Q$ beam-selection logits are produced by a linear classification layer:
\begin{equation}
\mathbf{o}_{n}^{(IDG)}
=
\mathbf{W}_{IDG,3}
\mathbf{u}_{n,IDG}^{(2)}
+
\mathbf{b}_{IDG,3} \in\mathbb{R}^{32},
\label{eq:final_logits}
\end{equation}

No GELU activation is applied to the final logit vector. The nonlinear GELU transformations are confined to the learned hidden representations, whereas the final linear logits are supplied directly to the softmax operation. The final beam-probability vector is
\begin{equation}
\widehat{\boldsymbol{\pi}}_n
=
\operatorname{softmax}
\left(
\mathbf{o}_{n}^{(IDG)}
\right).
\label{eq:final_probability}
\end{equation}

Equivalently, $\pi_{n,q} = \frac{ \exp\!\left(o_{n,q}^{(IDG)}\right) }{\sum_{\ell=0}^{Q-1} \exp\!\left(o_{n,\ell}^{(IDG)}\right)}$. 

The final beam prediction is:
\begin{equation}
\hat q_n
=
\underset{q\in\{0,\ldots,Q-1\}}
{\arg\max}
\;
\pi_{n,q}.
\label{eq:final_prediction}
\end{equation}

The nonlinear activation and normalization operations in \eqref{eq:image_projection}--\eqref{eq:idg_fusion_2} ensure that successive affine transformations do not collapse into a single linear mapping.

% =========================================================
\subsection{Staged Training Objective} \label{subsec:proposed_multimodal}
% =========================================================

The proposed model is trained progressively so that the contribution of each representation branch can be learned before the final geometry-guided fusion stage. For an auxiliary branch $m$, the categorical cross-entropy over a minibatch $\mathcal{I}_B$ is
\begin{equation}
\mathcal{L}_{\mathrm{CE}}^{(m)}
=
-\frac{1}{|\mathcal{I}_B|}
\sum_{n\in\mathcal{I}_B}
\log
\pi_{n,q_n^\star}^{(m)}.
\label{eq:aux_ce}
\end{equation}

The synthetic-data and image branches are first trained with their respective temporary classification heads. Their learned feature extractors are then retained while the temporary classifiers are discarded. The image--synthetic fusion head is subsequently trained using the frozen pretrained branch representations. Finally, the geometric--kinematic encoder and final fusion head are trained using the intermediate image-synthetic representation. During staged training, parameters belonging to previously trained stages remain frozen while the active stage is optimized. For the final stage, label smoothing is applied. For smoothing factor $\epsilon\in[0,1)$, the smoothed target for class $q$ is $t_{n,q} = (1-\epsilon)\mathbf{1}(q=q_n^\star) + \frac{\epsilon}{Q}, ~\epsilon=0.03$. The final label-smoothed cross-entropy loss is
\begin{equation}
\mathcal{L}_{\mathrm{LS}}
=
-\frac{1}{|\mathcal{I}_B|}
\sum_{n\in\mathcal{I}_B}
\sum_{q=0}^{Q-1}
t_{n,q}
\log
\pi_{n,q}.
\label{eq:training_loss}
\end{equation}

For $\epsilon=0$, \eqref{eq:training_loss} reduces to conventional one-hot cross-entropy. AdamW is employed for optimization. During final multimodal training, an initial learning rate of $4\times10^{-4}$ and weight decay of $10^{-4}$ are used together with a cosine-annealing learning-rate schedule over 32 epochs. Gradient clipping with a maximum norm of $1.0$ is applied during the final stage to improve optimization stability. The parameter state saved at the prescribed final epoch of each stage's fixed budget is retained for subsequent performance reporting. The overall staged training problem is accordingly decomposed into the following optimization subproblems:

\noindent
\begin{minipage}{0.45\columnwidth}
\begin{equation}
\boldsymbol{\Theta}_D^{\star} = \arg\min_{\boldsymbol{\Theta}_D} \mathcal{L}_{\mathrm{CE}}^{(D)}
\label{eq:synthetic_training_problem}
\end{equation}
\end{minipage}\hfill
\begin{minipage}{0.45\columnwidth}
\begin{equation}
\boldsymbol{\Theta}_I^{\star} = \arg\min_{\boldsymbol{\Theta}_I} \mathcal{L}_{\mathrm{CE}}^{(I)}
\label{eq:image_training_problem}
\end{equation}
\end{minipage}
\vspace{-4pt} % Adjust this value to squeeze vertical space between rows
\noindent
\begin{minipage}{0.49\columnwidth}
\begin{equation}
\boldsymbol{\Theta}_{ID}^{\star} = \arg\min_{\boldsymbol{\Theta}_{ID}} \mathcal{L}_{\mathrm{CE}}^{(ID)}
\label{eq:id_training_problem}
\end{equation}
\end{minipage}\hfill
\begin{minipage}{0.49\columnwidth}
\begin{equation}
\boldsymbol{\Theta}_{IDG}^{\star} = \arg\min_{\boldsymbol{\Theta}_{IDG}} \mathcal{L}_{\mathrm{LS}}^{(IDG)}
\label{eq:idg_training_problem}
\end{equation}
\end{minipage}

In \eqref{eq:id_training_problem}, the previously trained image and synthetic-data encoders remain frozen, and in the final optimization \eqref{eq:idg_training_problem}, the previously trained image--synthetic network is likewise frozen while only the geometry encoder and final classifier are trained. Algorithm~\ref{alg:staged_training} summarizes the staged optimization procedure. The optimization subproblems sequentially isolate $\boldsymbol{\Theta}_{\mathrm{D}}^{\star}$, $\boldsymbol{\Theta}_{\mathrm{I}}^{\star}$, $\boldsymbol{\Theta}_{\mathrm{ID}}^{\star}$, and finally $\boldsymbol{\Theta}_{\mathrm{IDG}}^{\star}$. For each test observation $n$, let $\mathcal{T}_{n,k} \subseteq \{0,\ldots,Q-1\}$ contain the indices of the $k$ largest predicted beam probabilities. The Top-$k$ beam-selection accuracy is defined as:
\begin{equation}
\operatorname{Acc}@k
=
\frac{1}{N_{\mathrm{te}}}
\sum_{n=1}^{N_{\mathrm{te}}}
\mathbf{1}
\left(
q_n^\star\in\mathcal{T}_{n,k}
\right),
\label{eq:topk_accuracy}
\end{equation}

Here, $N_{\mathrm{te}}$ denotes the number of test observations.

\begin{algorithm}[!t]
\caption{Staged Training of the Image--Synthetic--Geometry Beam Predictor}
\label{alg:staged_training}
\footnotesize
\begin{algorithmic}[1]

\Require Aligned training samples
$\{(\mathbf{I}_n,\mathbf{p}_n,q_n^\star)\}_{n=1}^{N_{\mathrm{tr}}}$;
learning rates $\eta_I,\eta_D,\eta_{ID},\eta_{IDG}$;
epoch budgets $T_I,T_D,T_{ID},T_{IDG}$;
mini-batch size $B$

\Ensure Trained parameters
$\boldsymbol{\Theta}_I^\star$,
$\boldsymbol{\Theta}_D^\star$,
$\boldsymbol{\Theta}_{ID}^\star$,
$\boldsymbol{\Theta}_{IDG}^\star$
\State \textbf{Preprocessing}
\State Align image and position records and construct
$q_n^\star$, $\mathbf{d}_n$, and $\mathbf{g}_n$
using \eqref{eq:measured_best_beam},
\eqref{eq:synthetic_vector}, and
\eqref{eq:geometry_vector}
\State Fit feature normalization statistics using the training partition only
\State \textbf{Stage 1: Synthetic-data encoder}
\For{$e=1$ to $T_D$}
    \State Compute $\mathbf{z}_{n,D}$ using
    \eqref{eq:synthetic_hidden}--\eqref{eq:synthetic_embedding}
    \State Minimize $\mathcal{L}_{\mathrm{CE}}^{(D)}$
    and update $\boldsymbol{\Theta}_D$ using AdamW with $\eta_D$
\EndFor
\State Retain $\boldsymbol{\Theta}_D^\star$
\State \textbf{Stage 2: Image encoder}
\For{$e=1$ to $T_I$}
    \State Compute $\mathbf{z}_{n,I}$ using
    \eqref{eq:image_backbone}--\eqref{eq:image_projection}
    \State Minimize $\mathcal{L}_{\mathrm{CE}}^{(I)}$
    and update $\boldsymbol{\Theta}_I$ using AdamW with $\eta_I$
\EndFor
\State Retain $\boldsymbol{\Theta}_I^\star$
\State \textbf{Stage 3: Image--Synthetic fusion}
\State Freeze $\boldsymbol{\Theta}_I^\star$ and
$\boldsymbol{\Theta}_D^\star$
\For{$e=1$ to $T_{ID}$}
    \State Form
    $\mathbf{c}_{n,ID}=[\mathbf{z}_{n,I};\mathbf{z}_{n,D}]$
    using \eqref{eq:image_synthetic_concat}
    \State Compute $\mathbf{z}_{n,ID}$ using
    \eqref{eq:id_fusion_1}--\eqref{eq:id_fusion_2}
    \State Minimize $\mathcal{L}_{\mathrm{CE}}^{(ID)}$
    and update $\boldsymbol{\Theta}_{ID}$ using AdamW with $\eta_{ID}$
\EndFor
\State Retain $\boldsymbol{\Theta}_{ID}^\star$
\State \textbf{Stage 4: Geometry-guided fusion}
\State Freeze $\boldsymbol{\Theta}_I^\star$,
$\boldsymbol{\Theta}_D^\star$, and
$\boldsymbol{\Theta}_{ID}^\star$
\For{$e=1$ to $T_{IDG}$}
    \State Compute $\mathbf{z}_{n,G}$ using
    \eqref{eq:geometry_hidden}--\eqref{eq:geometry_embedding}
    \State Form
    $\mathbf{c}_{n,IDG}=[\mathbf{z}_{n,ID};\mathbf{z}_{n,G}]$
    using \eqref{eq:idg_concat}
    \State Compute $\mathbf{o}_{n}^{(IDG)}$ using
    \eqref{eq:idg_fusion_1}--\eqref{eq:final_logits}
    \State Minimize $\mathcal{L}_{\mathrm{LS}}$ using
    \eqref{eq:training_loss}
    and update $\boldsymbol{\Theta}_{IDG}$ using AdamW with $\eta_{IDG}$
\EndFor

\State \Return
$\boldsymbol{\Theta}_I^\star$,
$\boldsymbol{\Theta}_D^\star$,
$\boldsymbol{\Theta}_{ID}^\star$,
$\boldsymbol{\Theta}_{IDG}^\star$

\end{algorithmic}
\end{algorithm}
\subsection{Inference-Time Beam Recommendation}
\label{subsec:inference_recommendation}

Algorithm~\ref{alg:inference_recommendation} describes how the trained staged model from Algorithm~\ref{alg:staged_training} is deployed at inference time. Given a new synchronized image and position observation, it applies the identical preprocessing pipeline used during training and evaluation, propagates the observation through each frozen stage of the network in the same order established during staged training, and returns an ordered Top-$k$ beam shortlist together with confidence scores, without requiring measured instantaneous CSI. 

In a deployed setting, the receiver subjects every new UAV observation to the same deterministic preprocessing pipeline used during evaluation. The RGB frame is transformed by the fixed evaluation image pipeline, while the current and previous position tuples are used to construct the synthetic position-derived vector $\mathbf{d}_n$ and the geometric--kinematic vector $\mathbf{g}_n$. The fitted normalization statistics are reused unchanged, so no test-set statistics, future trajectory samples, or measured instantaneous CSI are required. The trained classifier output logits $\mathbf{o}_n^{(IDG)}$, and the softmax values $\widehat{\boldsymbol{\pi}}_n$ are interpreted as beam-ranking scores rather than calibrated link-quality estimates. 

\begin{algorithm}[!t] \caption{Inference-Time Top-$k$ Beam Recommendation} \label{alg:inference_recommendation} \footnotesize \begin{algorithmic}[1]  \Require RGB frame $\mathbf{I}_n$, current position $(x_n,y_n)$, previous position $(x_{n-1},y_{n-1})$, trained parameters $\boldsymbol{\Theta}_I^\star,\boldsymbol{\Theta}_D^\star,\boldsymbol{\Theta}_{ID}^\star, \boldsymbol{\Theta}_{IDG}^\star$, fitted normalization statistics, shortlist size $k$  \Ensure Ordered Top-$k$ beam set $\mathcal{T}_{n,k}$  \State Apply fixed image preprocessing to $\mathbf{I}_n$ \State Construct $\mathbf{d}_n$ and $\mathbf{g}_n$ using \eqref{eq:synthetic_vector} and \eqref{eq:geometry_vector}; normalize $\mathbf{d}_n$ and standardize $\mathbf{g}_n$ using fitted training statistics  \Statex \textbf{Representation extraction} \State Compute $\mathbf{z}_{n,I}$ using \eqref{eq:image_backbone}--\eqref{eq:image_projection}, $\mathbf{z}_{n,D}$ using \eqref{eq:synthetic_hidden}--\eqref{eq:synthetic_embedding}, and $\mathbf{z}_{n,G}$ using \eqref{eq:geometry_hidden}--\eqref{eq:geometry_embedding}  \Statex \textbf{Staged fusion} \State Form $\mathbf{c}_{n,ID}=[\mathbf{z}_{n,I};\mathbf{z}_{n,D}]$ and compute $\mathbf{z}_{n,ID}$ using \eqref{eq:image_synthetic_concat}--\eqref{eq:id_fusion_2} \State Form $\mathbf{c}_{n,IDG}=[\mathbf{z}_{n,ID};\mathbf{z}_{n,G}]$ and compute $\mathbf{o}_{n}^{(IDG)}$ using \eqref{eq:idg_concat}--\eqref{eq:final_logits}  \Statex \textbf{Beam ranking} \State Compute $\widehat{\boldsymbol{\pi}}_n=\operatorname{softmax} (\mathbf{o}_{n}^{(IDG)})$ using \eqref{eq:final_probability}; sort beam indices by decreasing probability and retain the first $k$ as $\mathcal{T}_{n,k}$  \State \Return $\mathcal{T}_{n,k}$  \end{algorithmic} \end{algorithm}

The receiver first probes the beam with the highest rank. If the measured received power or the link-quality indicator is below the required threshold, the remaining beams in $\mathcal{T}_{n,k}$ provide a compact refinement set. This links the accuracy reported by Top-$k$ with the practical objective of reducing the number of beam trials before radio confirmation.

This fallback protocol gives operational meaning to the accuracy of Top-1, Top-3, and Top-5. Top-1 measures whether the first recommended beam is correct, whereas Top-$k$ measures whether the correct beam remains within a short candidate list. Thus, even when the absolute Top-1 prediction is missed, a high Top-$k$ value indicates that the correct beam is likely to be reachable from a small candidate set; this is evidence of candidate set compression under the present offline evaluation, not a demonstrated reduction in over-the-air beam-training trials, radio confirmation time, or protocol overhead, none of which were measured in this study.

\section{Experimental Setup}
\label{sec:experimental_setup}

% \subsection{Reproducibility and Data Protocol}
% All experiments were executed in Python with PyTorch and the \texttt{timm} implementation of EfficientNet on a Colab-hosted, CUDA-enabled GPU runtime. The pseudorandom seed was fixed at 42. The Scenario-23 dataset comprised 11,387 aligned samples, which were partitioned using a stratified 80/20 split with a fixed random seed of 42, yielding 9,109 samples for model fitting and 2,278 samples for evaluation. The same training partition was used for all four trained models (synthetic-only, image-only, image--synthetic, and full fusion). Mini-batches contained 16 observations.

\subsection{Reproducibility and Data Protocol}
\label{subsec:reproducibility}
All experiments were conducted on a reconstructed version of DeepSense~6G Scenario~23 that contained 11,387 aligned observations. The available Scenario~23 data subsets were combined to recover the complete set of synchronized image, position, and beam-sweep records, after which a fixed stratified 80/20 sample-level partition was generated using a pseudorandom seed of 42. This produced 9,109 samples for training and 2,278 samples for evaluation. The same partition was used for all model configurations to ensure a consistent basis for comparison.

All preprocessing operations were fitted using the training subset only. In particular, geometric--kinematic standardization parameters and synthetic-feature min--max normalization parameters were estimated from the training subset and subsequently applied unchanged to the evaluation subset. Stochastic image augmentation was restricted to the training subset, while evaluation used deterministic preprocessing. The same target construction, feature transformations, normalization procedures, and partition were applied across all experiments. Unless otherwise stated, the mini-batch size was 16 and the pseudorandom seed was fixed at 42. Model selection and final evaluation follow the protocol described in the corresponding experimental subsections.

To characterize temporal structure in the evaluation partition, we quantified the proximity between training and evaluation samples. For each evaluation sample, the minimum acquisition-timestamp distance to any training sample was computed. Scenario~23 was recorded at approximately 100--140~ms intervals. Consequently, 88.7\% of evaluation samples (1,768/2,278) have a training sample within 150~ms, while 99.9\% (2,275/2,278) have one within 1~s. Samples within 150~ms achieved 87.56\% Top-1 accuracy, compared with 85.69\% for the remaining 510 samples farther than 150~ms, a difference of 1.87 percentage points. The modest difference indicates that temporal proximity does not produce a large accuracy shift within this partition.

\subsection{Optimization Protocol}
Optimization followed the staged decomposition in Algorithm~\ref{alg:staged_training}, with a mini-batch size of 16 throughout. The synthetic-data branch was trained first for 100 epochs using AdamW with a learning rate of $3\times10^{-4}$ and weight decay of $10^{-3}$. The image branch was subsequently adapted for 20 epochs using AdamW, with learning rates of $5\times10^{-5}$ for the EfficientNet-B3 backbone and $3\times10^{-4}$ for its task head. The pretrained synthetic-data and image representations were then frozen while the image--synthetic fusion head was optimized for 20 epochs using AdamW with a learning rate of $3\times10^{-4}$ and weight decay of $10^{-4}$. Finally, explicit geometric--kinematic representation was introduced and the geometry-guided fusion stage was trained for 32 epochs using AdamW with a learning rate of $4\times10^{-4}$, weight decay of $10^{-4}$, cosine annealing, and gradient-norm clipping at 1.0.

The same reconstructed Scenario~23 partition was used for all four model configurations to ensure a consistent basis for comparison. The staged procedure isolates the contribution of each fusion stage by retaining previously learned representations while optimizing the newly introduced components. For every stage, the checkpoint saved at the prescribed final epoch of its fixed budget is retained; the test partition is not accessed during training or for checkpoint selection, and is evaluated exactly once after each stage's fixed epoch budget completes.

\section{Results and Discussion}
\label{sec:results}

Each evaluated checkpoint in this section corresponds to a specific stage
of the staged training procedure in Section~\ref{subsec:proposed_multimodal}
and Algorithm~\ref{alg:staged_training}. Synthetic only reports
$\boldsymbol{\Theta}_D^{\star}$ evaluated through the synthetic-data branch
alone; Image only reports $\boldsymbol{\Theta}_I^{\star}$ evaluated
through the image branch alone; Image + Synthetic reports the fused
embedding $\mathbf{z}_{n,ID}$ produced by $\boldsymbol{\Theta}_{ID}^{\star}$;
and Full fusion reports the complete geometry-guided pipeline
$\widehat{\boldsymbol{\pi}}_n$ produced by $\boldsymbol{\Theta}_{IDG}^{\star}$,
consistent with the inference procedure in
Algorithm~\ref{alg:inference_recommendation}.

\subsection{Ablation of the Input Representations}
\label{subsec:ablation}

Table~\ref{tab:ablation_results} compares four independently stored
checkpoints on the same fixed test partition. The Synthetic-only
branch obtains 60.71\% Top-1 accuracy, demonstrating that the deterministic
position-derived basis $\mathbf{d}_n$ is strongly predictive of the beam
labels along this trajectory. This association should not be interpreted
as recovery of a measured propagation channel. Image-only inference reaches
84.28\%, exceeding Synthetic-only inference by 23.57 percentage points and
establishing visual context as the stronger standalone representation under
the present split. The ablation therefore separates two claims that are
often conflated: structured position lifting carries useful predictive
information, but scene appearance supplies most of the unimodal
discriminative power.

\begin{table}[!t]
\caption{Trained component ablation on the fixed test partition.}
\label{tab:ablation_results}
\centering
\scriptsize
\setlength{\tabcolsep}{1.8pt}
\renewcommand{\arraystretch}{1.12}
\begin{tabular}{@{}llcccc@{}}
\hline
\textbf{Variant} & \textbf{Omitted} &
\textbf{Top-1 (\%)} & \textbf{Drop} &
\textbf{Top-3 (\%)} & \textbf{Top-5 (\%)} \\
\hline
Synthetic only
& RGB
& 60.71 [58.69, 62.70]
& 26.43
& 91.44
& 97.50 \\

Image only
& Synthetic
& 84.28 [82.73, 85.72]
& 2.86
& 98.73
& 99.34 \\

Image + Synthetic
& Geometry
&{86.92 [85.47, 88.24]}
& 0.22
& 99.03
& 99.56 \\

Full fusion
& None
& \textbf{87.14 [85.70, 88.45]}
& \textbf{0.00}
& 99.12
& 99.60 \\
\hline
\end{tabular}

\vspace{2pt}
\parbox{\columnwidth}{%
\footnotesize
\textit{Note:} Values in brackets are 95\% Wilson score confidence
intervals over the 2278-sample test partition. Drop is reported
relative to the highest measured configuration (Full fusion, 87.14\%
Top-1).
}
\end{table}

\begin{figure}[htbp]
    \centering
    \includegraphics[width=.9\columnwidth]{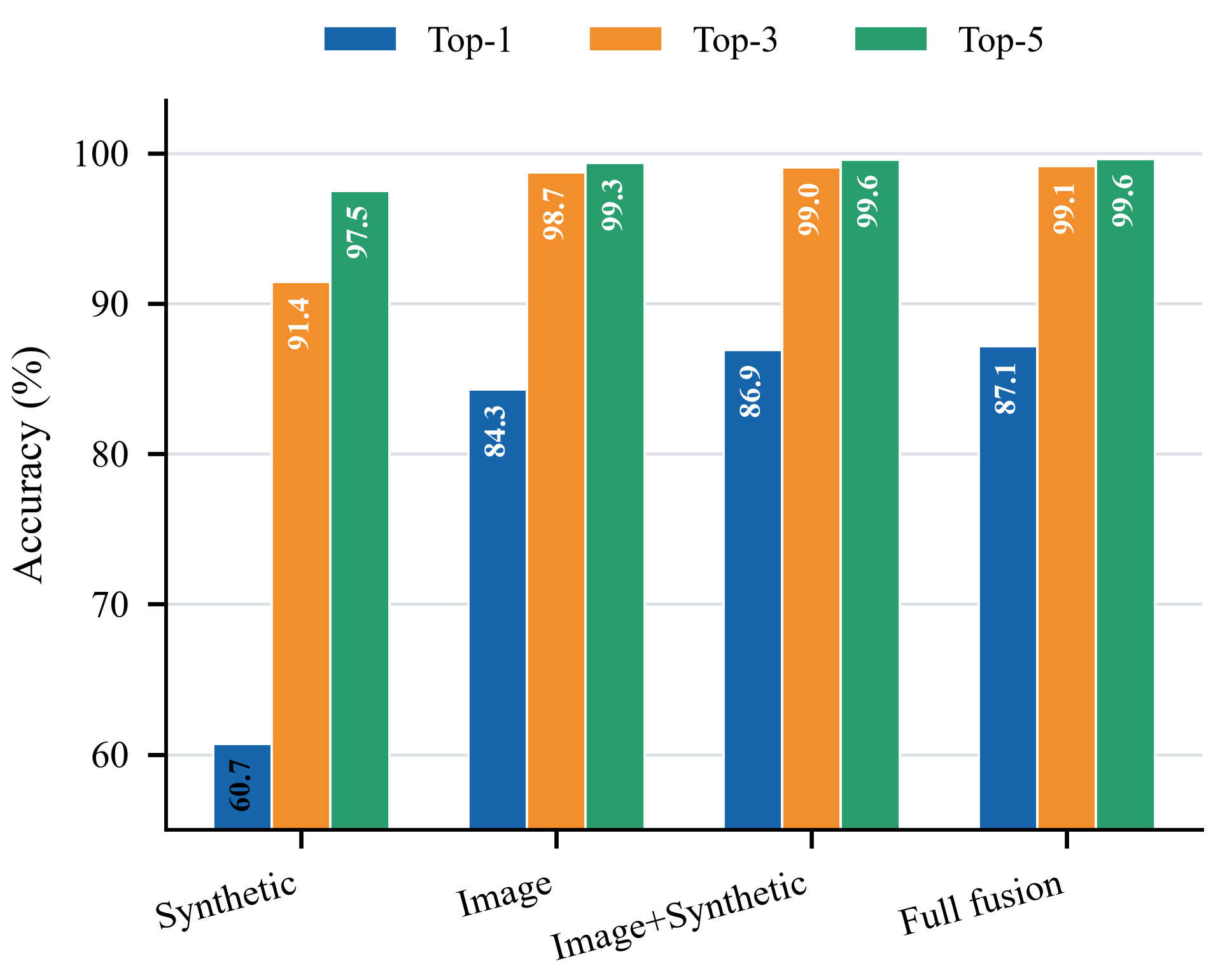}
    \caption{Ablation of the four implemented input configurations. Full fusion gives the highest measured result across all three candidate-list sizes.}
    \label{fig:ablation_topk}
\end{figure}

The Full fusion checkpoint provides the highest measured ranking performance: 87.14\% Top-1 (95\% CI: [85.70, 88.45]), 99.12\% Top-3 (2258/2278), and 99.60\% Top-5 (2269/2278) accuracy (Fig.~\ref{fig:ablation_topk}), ahead of Image + Synthetic at 86.92\% Top-1 (95\% CI: [85.47, 88.24]), 99.03\% Top-3 (2256/2278), and 99.56\% Top-5 (2268/2278). Relative to image-only inference, Image + Synthetic improves Top-1 accuracy by 2.64 percentage points, while Top-3 and Top-5 improve by 0.30 and 0.22 points, respectively. The near-saturated Top-3 and Top-5 values show that the remaining uncertainty is concentrated within a short ranked candidate list, consistent with the shortlist $\mathcal{T}_{n,k}$ returned by Algorithm~\ref{alg:inference_recommendation}.

Adding the explicit Geometric--Kinematic branch yields the highest measured performance, increasing Top-1 accuracy from 86.92\% to 87.14\% and Top-3 accuracy from 99.03\% to 99.12\%. The corresponding Top-5 accuracy increases from 99.56\% to 99.60\%. The incremental gain indicates that the principal performance improvement is obtained through image--synthetic fusion, while the explicit geometry branch provides an additional refinement to the multimodal representation.

Table~\ref{tab:feature_suppression} complements retrained-model ablation
with post-hoc interventions applied to the selected image--synthetic
checkpoint. The distinction is essential. Ablation compares models
optimized with different input sets; intervention probes the response of
one fixed decision function to inputs that may lie outside its training
support. Relative to the clean 86.92\% Top-1 result, zeroing all 11
synthetic coordinates produces a 78.05-point loss, while replacing the
image with a black frame produces a 41.97-point loss. The 2.46-point
decrease under the prescribed deterministic synthetic-feature shift (DSFS) establishes
tolerance only to that synthetic transformation. Group-level intervention is appropriate because
the synthetic coordinates in \eqref{eq:synthetic_vector} are correlated
deterministic transformations of a common $(x_n,y_n)$ record and therefore
do not constitute independent sensing variables.

\begin{table}[t]
\centering
\caption{Feature-group suppression sensitivity of the image--synthetic predictor.}
\label{tab:feature_suppression}
\begin{tabular}{lccc}
\hline
\textbf{Suppressed or Degraded Cue} &
\textbf{Top-1} &
\textbf{Drop} &
\textbf{Retained} \\
\hline
None (clean input)                         & 86.92 & 0.00  & 100.0\% \\
Visual content (black frame)               & 44.60 & 41.97 & 51.5\%  \\
Visual intensity (25\% brightness)         & 54.08 & 32.49 & 62.5\%  \\
Synthetic vector (all 11 entries zeroed)   & 8.52  & 78.05 & 9.8\%   \\
Synthetic structure (DSFS)                 & 84.11 & 2.46  & 97.2\%  \\
Synthetic precision (noise, $\sigma=0.1$)  & 64.53 & 22.04 & 74.5\%  \\
\hline
\end{tabular}
\end{table}

\begin{figure}[htbp]
    \centering
    \includegraphics[width=\columnwidth]{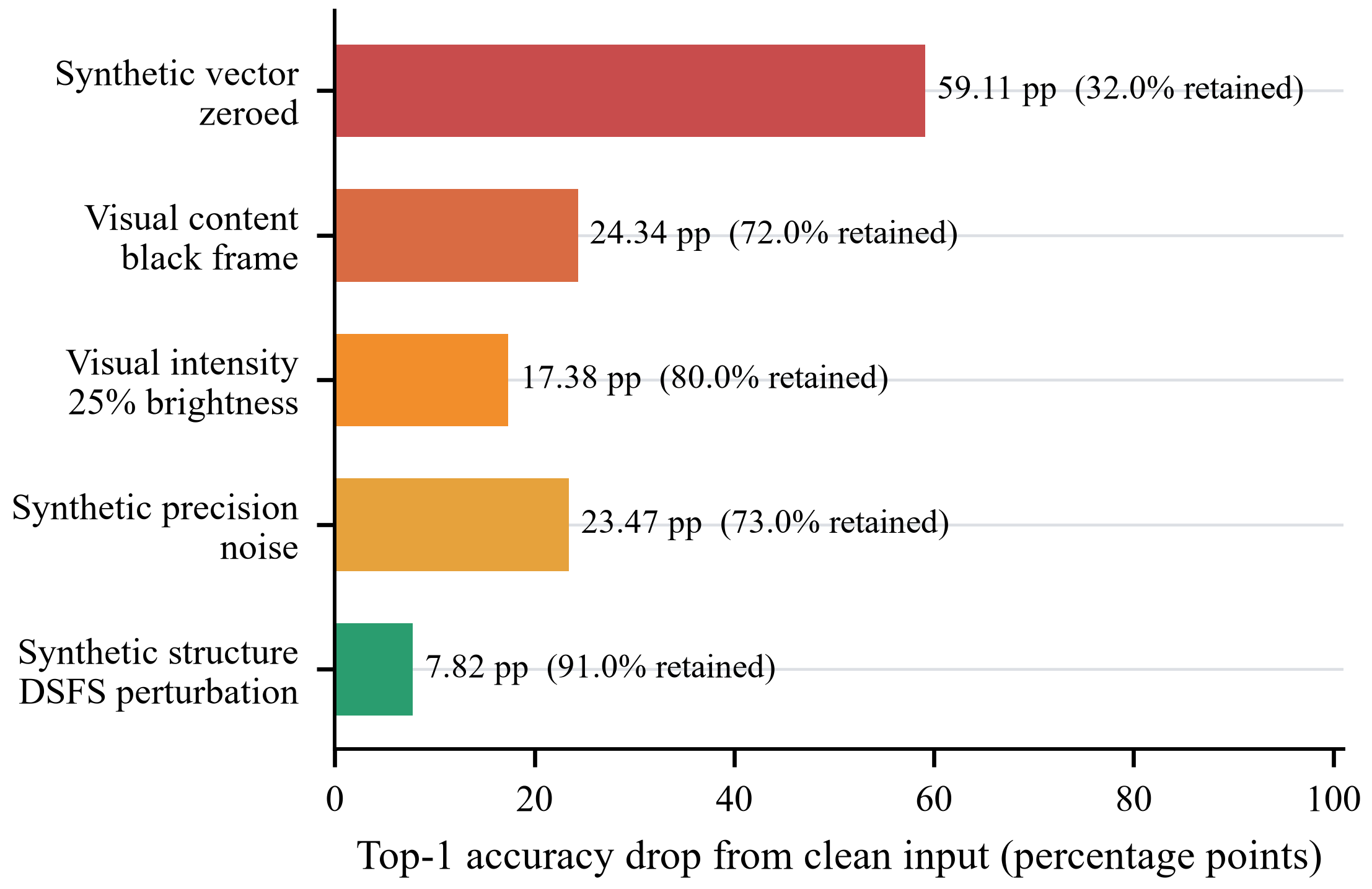}
    \caption{Grouped input-intervention sensitivity of the selected image-synthetic predictor. Bar length denotes the Top-1 loss relative to the clean 86.92\% result; parenthetical values report the retained fraction of clean accuracy.}
    \label{fig:feature_suppression_drop}
\end{figure}

Figure~\ref{fig:feature_suppression_drop} reveals a marked asymmetry in the
sensitivity of the joint classifier. Complete synthetic suppression is
almost twice as damaging as image suppression, whereas the specified
DSFS retains 97.2\% of clean-input accuracy. These
magnitudes characterize the local behavior of the trained fusion interface
under the chosen interventions; they do not, by themselves, order the
intrinsic information content of the two modalities.

That qualification is supported by the independently trained image branch,
which remains accurate without synthetic input. The catastrophic response
to an all-zero synthetic vector instead indicates that the joint decision
surface has encoded a strong dependence on the expected synthetic
manifold and lacks an explicit mechanism for recognizing modality absence.
In other words, the failure is architectural and distributional: an
invalid input is processed as though it were a valid low-valued
observation.

\subsection{Controlled Input-Stress Evaluation}
\label{subsec:input_stress}

Nominal accuracy alone does not characterize multimodal behavior when an input branch is degraded. Table~\ref{tab:stress_results}
therefore evaluates deterministic test-time perturbations spanning two
qualitatively different failure families: visual degradation (a black
frame and 25\% brightness) and synthetic degradation (complete
suppression, the prescribed deterministic synthetic-feature shift, and additive
Gaussian noise with standard deviation 0.1). These experiments are
controlled diagnostics of the learned decision function. They should not
be interpreted as empirical models of naturally occurring blockage, NLOS
propagation, or sensor failure.

\begin{table}[h]
\caption{Top-1 Accuracy Under Controlled Test-Time Perturbations}
\label{tab:stress_results}
\centering
\scriptsize
\setlength{\tabcolsep}{2.4pt}
\renewcommand{\arraystretch}{1.10}
\begin{tabular}{lccc}
\hline
\textbf{Condition} & \textbf{Synthetic} & \textbf{Image} & \textbf{Image+Synthetic} \\
\hline
Clean input & 60.71 & 84.28 & \textbf{86.92} \\
Black image & \textbf{60.71} & 23.18 & 44.60 \\
25\% image brightness & \textbf{60.71} & 25.37 & 54.08 \\
Zeroed synthetic vector & 4.83 & \textbf{84.28} & 8.52 \\
DSFS perturbation & 31.43 & \textbf{84.28} & 84.11 \\
Synthetic noise, $\sigma=0.1$ & 17.56 & \textbf{84.28} & 64.53 \\
\hline
\end{tabular}
\\[2pt]
\footnotesize{DSFS: deterministic synthetic-feature shift (Section~\ref{subsec:input_stress}).}
\end{table}

The controlled interventions reveal a clear complementarity between the two branches under visual degradation. A black frame reduces image-only accuracy to 23.18\%, while image--synthetic fusion retains 44.60\%; at 25\% brightness, the corresponding values are 25.37\% and 54.08\%. The position-derived branch therefore supplies useful predictive information when visual content is degraded, although static fusion does not recover the full Synthetic-only baseline when the image becomes uninformative.

The reciprocal intervention is substantially more severe. Suppressing the synthetic vector reduces the fused accuracy from 86.92\% to 8.52\%, despite the independently trained image branch retaining 84.28\%. The Gaussian perturbation of the synthetic representation similarly reduces the fusion to 64.53\%, whereas the prescribed deterministic shift in synthetic-features produces a smaller 2.46-point decrease to 84.11\%. Together, these results show that the clean-input fusion head strongly depends on the expected synthetic representation and does not automatically compensate for severe synthetic-input degradation.

\begin{figure}[htbp]
    \centering
    \includegraphics[width=\columnwidth]{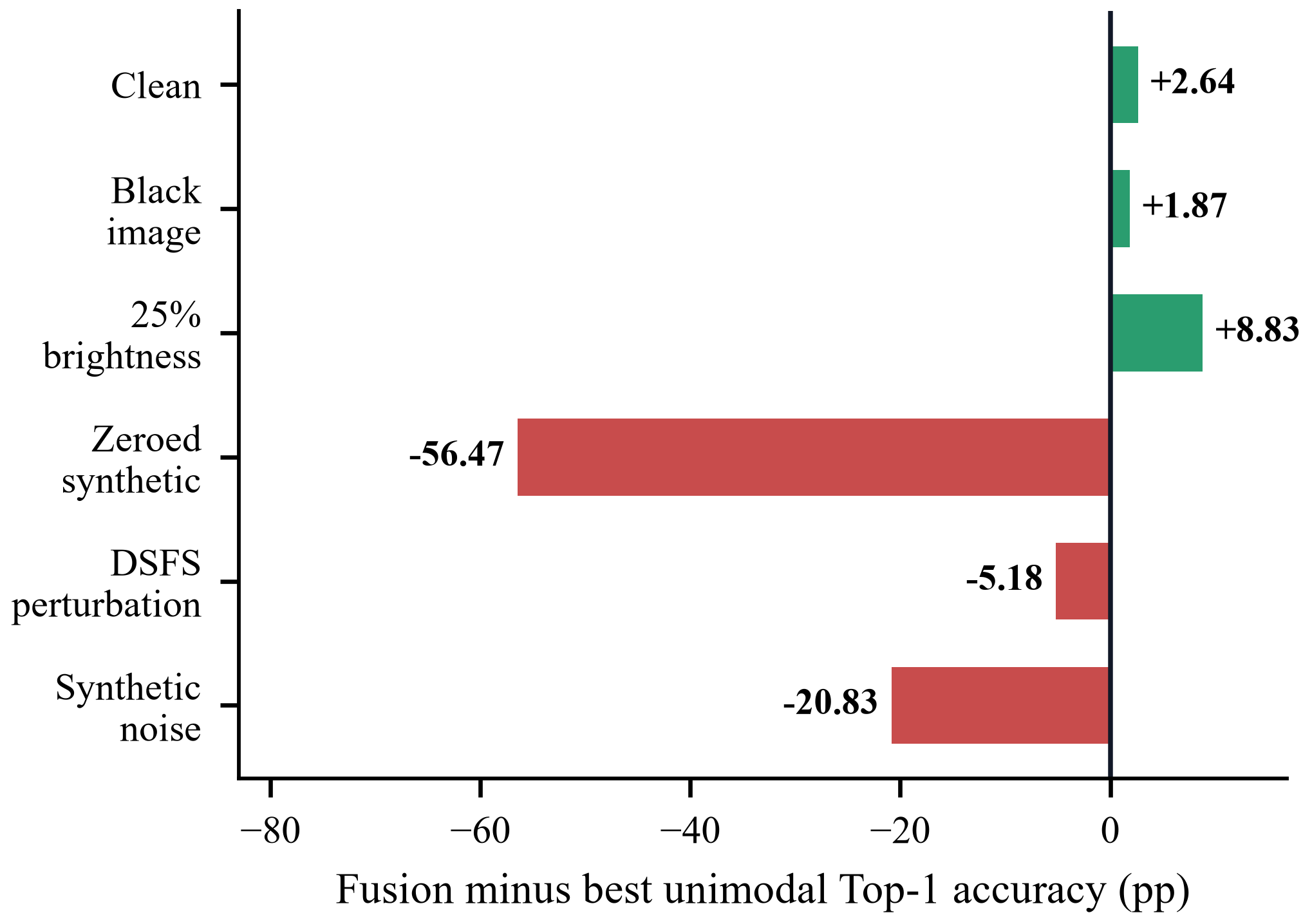}
    \caption{Fusion margin relative to the stronger independently trained unimodal fallback for each input condition. Positive values favor fusion; negative values quantify the penalty of static fusion.}
    \label{fig:fusion_fallback_gap}
\end{figure}

Figure~\ref{fig:fusion_fallback_gap} compares the fusion with the independently trained unimodal predictor associated with each condition. Fusion retains its 2.64-point advantage on clean inputs, while the evaluated perturbations reverse this ordering. The fallback deficit is 15.72 points for a black frame, 6.24 points at
25\% brightness, and 76.91 points when the synthetic vector is
suppressed. Even the comparatively mild DSFS leaves
the fusion 1.32 points below the unchanged image-only branch. Therefore, static feature
concatenation cannot be assumed to route decisions toward the
healthier modality. The observed failure pattern motivates three concrete
mechanisms for subsequent evaluation: explicit modality-presence
indicators, branch-wise reliability or uncertainty estimates, and
stochastic modality masking during training.

\subsection{Explainable Multimodal Attribution Analysis}
\label{subsec:xai}

To quantify how the measured joint performance is distributed between input groups, we used Shapley values at the group-level of Top-1 accuracy over the entire test partition. Let $A(S)$ denote the accuracy of Top-1 when only the subset $S$ is available, let $\mathcal{M}=\{I,D\}$ represent the image and synthetic-data groups consistent with the branch subscripts used throughout Section~\ref{subsec:proposed_multimodal}, and define the empty coalition by selecting the random $Q$-class, $A(\varnothing)=100/Q=100/32=3.125\%$. The attribution assigned to group
$m$ is
\begin{equation}
 \phi_m=\sum_{S\subseteq\mathcal{M}\setminus\{m\}}
 \frac{|S|!(|\mathcal{M}|-|S|-1)!}{|\mathcal{M}|!}
 \left[A(S\cup\{m\})-A(S)\right].
 \label{eq:group_shapley}
\end{equation}
Unlike pixel-level saliency, this construction attributes the measured utility of complete sensing branches and allocates their interaction according to the Shapley axioms. It is reproducible from full-partition coalition scores and does not depend on a selectively chosen image. Its scope is functional rather than causal: $\phi_m$ describes how prediction performance changes across the evaluated coalitions, not the physical effect of intervening in a sensor.

\begin{figure}[htbp]
    \centering
    \includegraphics[width=.85\columnwidth]{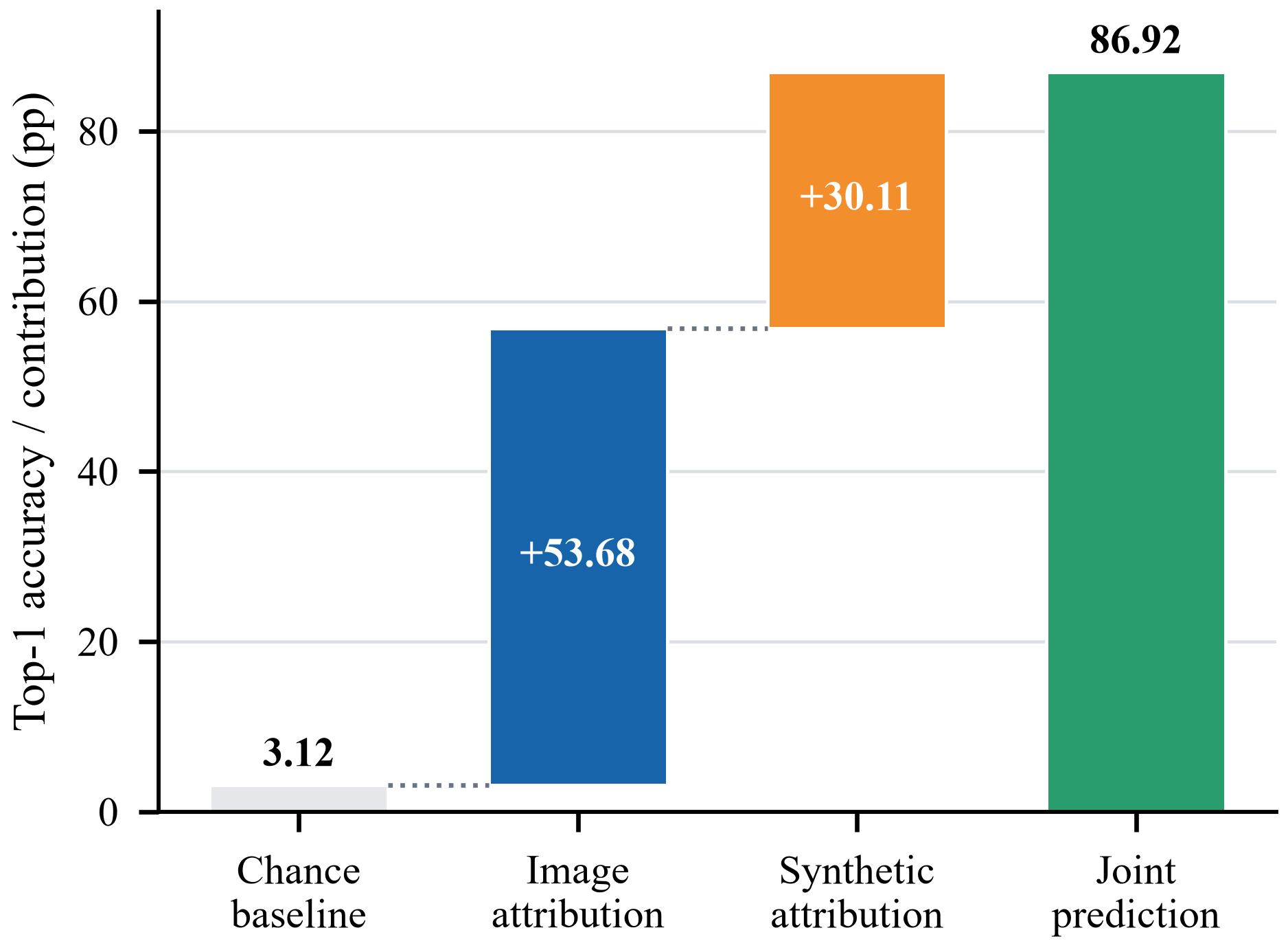}
    \caption{Clean-input coalition-performance decomposition of Top-1 accuracy across separately trained unimodal and fused checkpoints. Relative to the 3.125\% chance baseline, the image and synthetic-data groups contribute 53.68 and 30.11 percentage points, respectively, recovering the 86.92\% joint result.}
    \label{fig:xai_global_shapley}
\end{figure}

Figure~\ref{fig:xai_global_shapley} decomposes the 86.92\% joint result into interaction-aware allocations for each input group. 

Relative to the 3.125\% chance baseline, the image branch contributes 53.68 percentage points, and the synthetic-data branch contributes 30.11 percentage points. Thus, the vision and synthetic branches account for approximately 64.1\% and 35.9\% of the above-chance attributable performance, respectively. The decomposition is consistent with the clean-input ablation: visual context provides the dominant contribution, while the structured position-derived representation supplies a substantial complementary contribution.

This agreement between ablation and attribution is important because the two analyses answer different questions. Ablation measures the performance of separately optimized input configurations, whereas the  group Shapely analysis allocates the performance of the joint predictor across coalition values. Their consistent ordering strengthens the conclusion that the synthetic-data branch complements, but does not supersede, visual information on the clean partition. 

\begin{figure}[htbp]
    \centering
    \includegraphics[width=0.7\columnwidth]{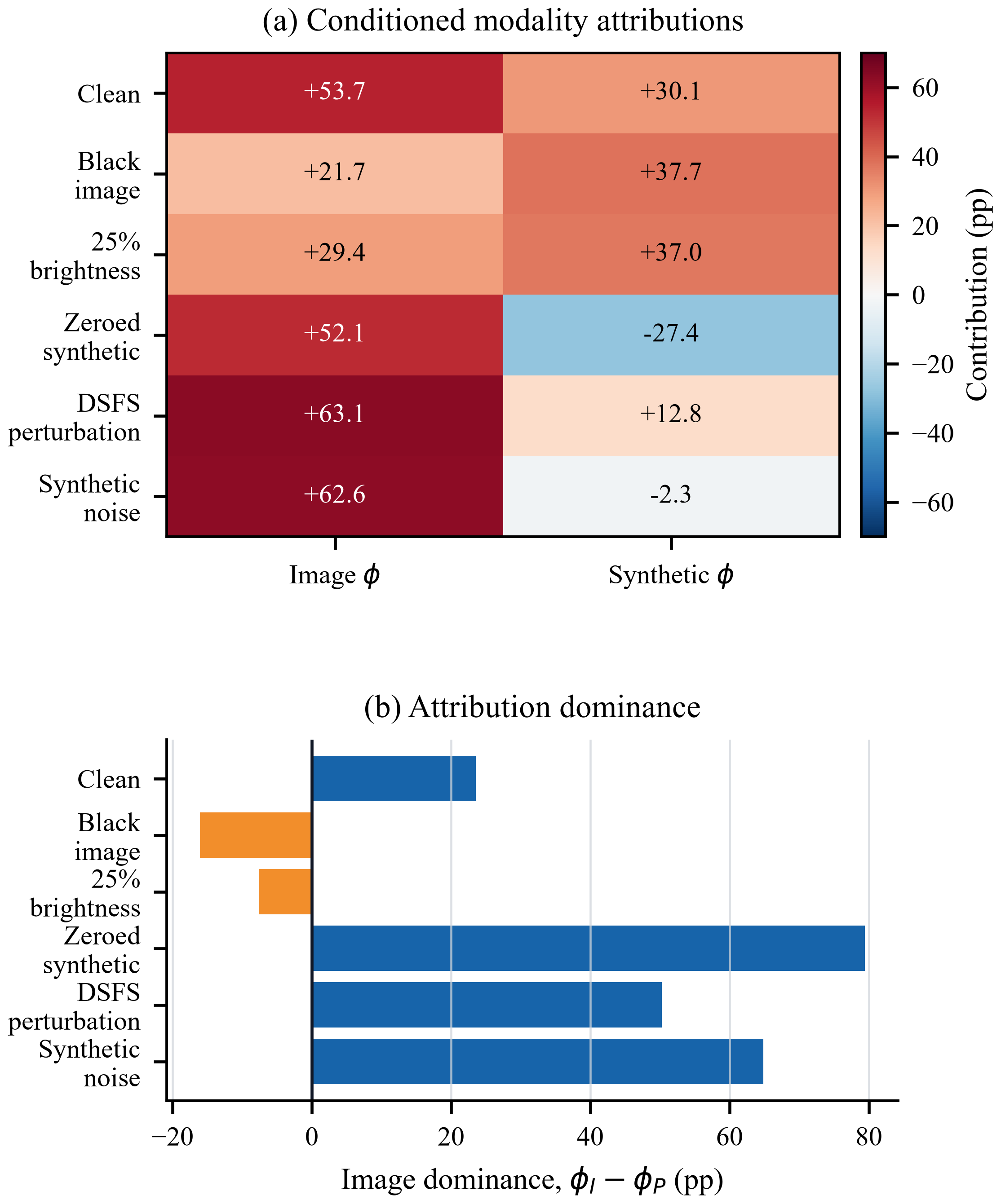}
    \caption{Corruption-conditioned coalition-performance diagnostics. (a) Signed modality contributions, where a negative value indicates that the corrupted branch reduces coalition accuracy. (b) Image-minus-synthetic attribution dominance for the corresponding conditions.}
    \label{fig:xai_conditioned_shapley}
\end{figure}

Figure~\ref{fig:xai_conditioned_shapley} exposes a structured redistribution of model reliance under degradation: visual attribution contracts sharply under image corruption, while synthetic-data attribution rises to compensate, and the reverse pattern holds under synthetic corruption. A suppressed synthetic vector receives a strongly negative attribution, meaning that the corrupted branch is not only ignored, but actively moves the joint predictor away from the decision supported by the intact image branch. This shift in attribution connects the aggregate stress-test losses to a failure mechanism at the modality-level and provides direct motivation for quality-gated or uncertainty-weighted fusion.

\section{Conclusion}
\label{sec:conclusion}

This study developed and evaluated a multimodal staged framework for current-beam prediction in real UAV mmWave communication using RGB imagery and position-derived representations from DeepSense~6G Scenario~23. The image--synthetic fusion model achieved 86.92\% Top-1, 99.03\% Top-3, and 99.56\% Top-5 accuracy, improving Top-1 performance by 2.64 percentage points over the image-only branch. Incorporating the explicit Geometric--Kinetic representation produced the highest measured performance, reaching 87.14\% Top-1, 99.12\% Top-3, and 99.60\% Top-5 accuracy, with a smaller incremental improvement over image--synthetic fusion. Controlled input interventions showed that the position-derived representation remains useful under visual degradation, while severe synthetic-input corruption can substantially degrade the fused predictor despite the strong standalone image branch. These results establish a practical position-assisted multimodal beam-prediction pipeline for the evaluated Scenario~23 and highlight the importance of modality-aware fusion for robust deployment. Future work will investigate reliability-aware modality gating, broader trajectory conditions, and deployment-oriented latency and resource measurements.


\begin{thebibliography}{33}
\raggedright

\bibitem{sabuj2023cuav}
S. R. Sabuj, S. Ahmed, and H.-S. Jo,
``Multiple CUAV-Enabled mMTC and URLLC Services: Review of Energy Efficiency and Latency Performance,''
\emph{IEEE Trans. Green Commun. Netw.}, vol. 7, no. 3, pp. 1369--1382, Sept. 2023,
doi: 10.1109/TGCN.2023.3281350.

\bibitem{charan2022drone}
G. Charan, A. Hredzak, C. Stoddard, B. Berrey, M. Seth, H. Nunez, and A. Alkhateeb,
``Towards real-world 6G drone communication: Position and camera aided beam prediction,''
in \emph{Proc. IEEE Global Commun. Conf. (GLOBECOM)}, 2022, pp. 2951--2956,
doi: 10.1109/GLOBECOM48099.2022.10000718.

\bibitem{alkhateeb2023deepsense}
A. Alkhateeb, G. Charan, T. Osman, A. Hredzak, J. Morais, U. Demirhan, and N. Srinivas,
``DeepSense 6G: A large-scale real-world multi-modal sensing and communication dataset,''
\emph{IEEE Commun. Mag.}, vol. 61, no. 9, pp. 122--128, Sept. 2023,
doi: 10.1109/MCOM.006.2200730.

\bibitem{morais2023position}
J. Morais, A. Behboodi, H. Pezeshki, and A. Alkhateeb,
``Position-aided beam prediction in the real world: How useful GPS locations actually are?''
in \emph{Proc. IEEE Int. Conf. Commun. (ICC)}, 2023, pp. 1824--1829,
doi: 10.1109/ICC45041.2023.10278998.

\bibitem{charan2022multimodal}
G. Charan, T. Osman, A. Hredzak, N. Thawdar, and A. Alkhateeb,
``Vision-Position Multi-Modal Beam Prediction Using Real Millimeter Wave Datasets,''
in \emph{Proc. IEEE Wireless Commun. Netw. Conf. (WCNC)}, 2022, pp. 2727--2731,
doi: 10.1109/WCNC51071.2022.9771835.

\bibitem{jiang2022vision}
S. Jiang and A. Alkhateeb,
``Computer vision aided beam tracking in a real-world millimeter wave deployment,''
in \emph{Proc. IEEE GLOBECOM Workshops}, 2022, pp. 142--147,
doi: 10.1109/GCWkshps56602.2022.10008648.

\bibitem{ahmad2023vision}
I. Ahmad, A. R. Khan, R. N. B. Rais, A. Zoha, M. A. Imran, and S. Hussain,
``Vision-assisted beam prediction for real world 6G drone communication,''
in \emph{Proc. IEEE 34th Annu. Int. Symp. Pers., Indoor Mobile Radio Commun. (PIMRC)}, 2023, pp. 1--7,
doi: 10.1109/PIMRC56721.2023.10294031.

\bibitem{jiang2023lidar}
S. Jiang, G. Charan, and A. Alkhateeb,
``LiDAR aided future beam prediction in real-world millimeter wave V2I communications,''
\emph{IEEE Wireless Commun. Lett.}, vol. 12, no. 2, pp. 212--216, Feb. 2023,
doi: 10.1109/LWC.2022.3219409.

\bibitem{luo2023radar}
H. Luo, U. Demirhan, and A. Alkhateeb,
``Millimeter wave V2V beam tracking using radar: Algorithms and real-world demonstration,''
in \emph{Proc. 31st Eur. Signal Process. Conf. (EUSIPCO)}, 2023, pp. 740--744,
doi: 10.23919/EUSIPCO58844.2023.10289752.

\bibitem{charan2025sensing}
G. Charan and A. Alkhateeb,
``Sensing-aided 6G drone communications: Real-world datasets and demonstration,''
\emph{IEEE Open J. Commun. Soc.}, vol. 6, pp. 8745--8774, 2025,
doi: 10.1109/OJCOMS.2025.3612320.

\bibitem{zarei2025lstm}
Z. Zarei, F. D. Tilahun, and C. G. Kang,
``LSTM-based beam prediction for 6G drone communications: Image vs. GPS-based sensing,''
in \emph{Proc. 30th Asia-Pacific Conf. Commun. (APCC)}, 2025,
doi: 10.23919/APCC64555.2025.11279910.

\bibitem{zarei2023seresnet}
Z. Zarei, F. D. Tilahun, and C. G. Kang,
``Vision-assisted beam prediction for UAV-enabled millimeter-wave communications using SE-ResNet50,''
in \emph{Proc. 14th Int. Conf. Inf. Commun. Technol. Converg. (ICTC)}, 2023, pp. 1659--1661,
doi: 10.1109/ICTC58733.2023.10392688.

\bibitem{yang2023semantics}
Y. Yang, F. Gao, X. Tao, G. Liu, and C. Pan,
``Environment semantics aided wireless communications: A case study of mmWave beam prediction and blockage prediction,''
\emph{IEEE J. Sel. Areas Commun.}, vol. 41, no. 7, pp. 2025--2040, Jul. 2023,
doi: 10.1109/JSAC.2023.3280966.

\bibitem{charan2025multicandidate}
G. Charan, M. Alrabeiah, T. Osman, and A. Alkhateeb,
``Camera based mmWave beam prediction: Towards multi-candidate real-world scenarios,''
\emph{IEEE Trans. Veh. Technol.}, vol. 74, no. 4, pp. 5897--5913, 2025,
doi: 10.1109/TVT.2024.3506948.

\bibitem{tan2019efficientnet}
M. Tan and Q. V. Le,
``EfficientNet: Rethinking model scaling for convolutional neural networks,''
in \emph{Proc. 36th Int. Conf. Mach. Learn. (ICML)}, vol. 97, pp. 6105--6114, 2019.

\bibitem{demirhan2022radarbeam}
U. Demirhan and A. Alkhateeb,
``Radar aided 6G beam prediction: Deep learning algorithms and real-world demonstration,''
in \emph{Proc. IEEE Wireless Commun. Netw. Conf. (WCNC)}, 2022, pp. 2655--2660,
doi: 10.1109/WCNC51071.2022.9771564.

\bibitem{ma2023dualband}
K. Ma, S. Du, H. Zou, W. Tian, Z. Wang, and S. Chen,
``Deep learning assisted mmWave beam prediction for heterogeneous networks: A dual-band fusion approach,''
\emph{IEEE Trans. Commun.}, vol. 71, no. 1, pp. 115--130, Jan. 2023,
doi: 10.1109/TCOMM.2022.3222345.

\bibitem{catak2022security}
F. O. Catak, M. Kuzlu, E. Catak, U. Cali, and D. Unal,
``Security concerns on machine learning solutions for 6G networks in mmWave beam prediction,''
\emph{Physical Commun.}, vol. 52, Art. no. 101626, 2022,
doi: 10.1016/j.phycom.2022.101626.

\bibitem{charan2022challenge}
G. Charan, U. Demirhan, J. Morais, A. Behboodi, H. Pezeshki, and A. Alkhateeb,
``Multi-modal beam prediction challenge 2022: Towards generalization,''
\emph{arXiv preprint arXiv:2209.07519}, 2022,
doi: 10.48550/arXiv.2209.07519.

\bibitem{li2023timedomain}
Q. Li \emph{et al.},
``Machine learning based time domain millimeter-wave beam prediction for 5G-Advanced and beyond: Design, analysis, and over-the-air experiments,''
\emph{IEEE J. Sel. Areas Commun.}, vol. 41, no. 6, pp. 1787--1809, Jun. 2023,
doi: 10.1109/JSAC.2023.3275613.

\bibitem{tian2023transformers}
Y. Tian, Q. Zhao, Z. E. A. Kherroubi, F. Boukhalfa, K. Wu, and F. Bader,
``Multimodal transformers for wireless communications: A case study in beam prediction,''
\emph{ITU J. Future Evol. Technol.}, vol. 4, no. 3, pp. 461--471, 2023,
doi: 10.52953/JWRA8095.

\bibitem{imran2023semantic}
S. Imran, G. Charan, and A. Alkhateeb,
``Environment semantic aided communication: A real world demonstration for beam prediction,''
in \emph{Proc. IEEE Int. Conf. Commun. Workshops (ICC Workshops)}, 2023, pp. 48--53,
doi: 10.1109/ICCWorkshops57953.2023.10283602.

\bibitem{khan2024lowcomplexity}
M. Q. Khan, A. Gaber, M. Parvini, P. Schulz, and G. Fettweis,
``A low-complexity machine learning design for mmWave beam prediction,''
\emph{IEEE Wireless Commun. Lett.}, vol. 13, no. 6, pp. 1551--1555, Jun. 2024,
doi: 10.1109/LWC.2024.3381447.

\bibitem{shi2024multimodal}
B. Shi, M. Li, M.-M. Zhao, M. Lei, and L. Li,
``Multimodal deep learning empowered millimeter-wave beam prediction,''
in \emph{Proc. IEEE 99th Veh. Technol. Conf. (VTC2024-Spring)}, 2024, pp. 1--6,
doi: 10.1109/VTC2024-Spring62846.2024.10683225.

\bibitem{xue2024survey}
Q. Xue \emph{et al.},
``A survey of beam management for mmWave and THz communications towards 6G,''
\emph{IEEE Commun. Surveys Tuts.}, vol. 26, no. 3, pp. 1520--1559, 2024,
doi: 10.1109/COMST.2024.3361991.

\bibitem{gouda2025position}
A. A. E. M. M. Gouda, E. K. I. Hamad, A. I. Hussein, M. M. Mabrook, and A. A. Donkol,
``Enhanced position-aided beam prediction using real-world data and enhanced-convolutional neural networks,''
\emph{IEEE Access}, vol. 13, pp. 74917--74929, 2025,
doi: 10.1109/ACCESS.2025.3563797.

\bibitem{nugroho2025gps}
V. A. Nugroho and B. M. Lee,
``GPS-aided deep learning for beam prediction and tracking in UAV mmWave communication,''
\emph{IEEE Access}, vol. 13, pp. 117065--117077, 2025,
doi: 10.1109/ACCESS.2025.3586594.

\end{thebibliography}
\end{document}